\documentclass[fleqn,usenatbib]{mnras}

\usepackage[T1]{fontenc}
\usepackage{graphicx}	% Including figure files
\usepackage{amsmath}	% Advanced maths commands
\usepackage{amssymb}	% Extra maths symbols
\usepackage{placeins}
\usepackage{newtxtext,newtxmath}
\newcommand{\LCDM}{$\Lambda$CDM }

\newcommand{\hpMpc}{$h \textrm{Mpc}^{-1}$}
\newcommand{\Mpcph}{$h^{-1} \textrm{Mpc}$}
\newcommand{\Euclid}{\textsc{Euclid}}
\newcommand{\LSST}{\textsc{LSST}}
\newcommand{\NGRST}{\textsc{NGRST}}
\newcommand{\FLASK}{\texttt{FLASK}}
\title[Constraining cosmological extensions with cosmic shear]{Testing extensions to $\Lambda$CDM on small scales with forthcoming cosmic shear surveys}

\author[S. G. Stafford et al.]{
Sam G. Stafford$^{1}$\thanks{E-mail: s.stafford@2014.ljmu.ac.uk}, Ian G. McCarthy$^{1}$\thanks{E-mail: i.g.mccarthy@ljmu.ac.uk}, Juliana Kwan$^{1}$, Shaun T. Brown$^1$, Andreea S. Font$^1$, \newauthor Andrew Robertson$^2$
\\
$^{1}$Astrophysics Research Institute, Liverpool John Moores University, 146 Brownlow Hill, Liverpool L3 5RF, UK \\
$^{2}$Institute for Computational Cosmology, Durham University, South Road, Durham DH1 3LE, UK\\
}

\date{Accepted XXX. Received YYY; in original form ZZZ}

\pubyear{2015}

\begin{document}
\label{firstpage}
\pagerange{\pageref{firstpage}--\pageref{lastpage}}
\maketitle

\begin{abstract}
We investigate the constraining power of forthcoming Stage-IV weak lensing surveys (\Euclid, \LSST, and \NGRST) for extensions to the $\Lambda$CDM model on small scales, via their impact on the cosmic shear power spectrum.  We use high-resolution cosmological simulations to calculate how warm dark matter (WDM), self-interacting dark matter (SIDM) and a running of the spectral index affect the non-linear matter power spectrum, $P(k)$, as a function of scale and redshift.  We evaluate the cosmological constraining power using synthetic weak lensing observations derived from these power spectra and that take into account the anticipated source densities, shape noise and cosmic variance errors of upcoming surveys. 
We show that upcoming Stage-IV surveys will be able to place useful, independent constraints on both WDM models (ruling out models with a particle mass of $\la0.5$ keV) and SIDM models (ruling out models with a velocity-independent cross-section of $\ga10$ cm$^2$ g$^{-1}$) through their effects on the small-scale cosmic shear power spectrum.  Similarly, they will be able to strongly constrain cosmologies with a running spectral index.  
Finally, we explore the error associated with the cosmic shear cross-spectrum between tomographic bins, finding that it can be significantly affected by Poisson noise (the standard assumption is that the Poisson noise cancels between tomographic bins).  We provide a new analytic form for the error on the cross-spectrum which accurately captures this effect.
\end{abstract}


%%%%%%%%%%%%%%%%%%%%%%%%%%%%%%%%%%%%%%%%%%%%%%%%%%

%%%%%%%%%%%%%%%%% BODY OF PAPER %%%%%%%%%%%%%%%%%%

\section{Introduction}

In the current concordance cosmological framework, termed the $\Lambda$CDM ($\Lambda$-cold dark matter) model, structure in the Universe forms hierarchically. The initial density perturbations laid down by inflation eventually grow large enough that they become gravitationally unstable and collapse to form low-mass `haloes'. 
These low-mass haloes merge to build up progressively larger systems, eventually culminating in the large-scale structure (LSS) that we observe today (see, e.g., \citealt{Davis1985}).  This theoretical picture has been immensely powerful in describing observations of our Universe, accurately reproducing the observed properties of the cosmic microwave background (see e.g. the recent results from \citealt{Planck2018}) as well as low-redshift probes such as baryon acoustic oscillations (\citealt{Eisensteien2005}) and redshift-space distortions (see e.g. \citealt{Alam2017}).  Given its success, this theoretical framework has come to be known as the `standard model of cosmology'.  

It is noteworthy that, while the $\Lambda$CDM model accurately describes many observables on large scales, there have been a number of recent mild tensions reported in the best-fit parameter values for certain cosmological parameters, including the Hubble constant (see \citealt{Verde2019} for a review) and the LSS parameter $S_8 \equiv \sigma_8 \sqrt{\Omega_m/0.3}$ (see, e.g., discussion in \citealt{McCarthy2018}) as derived from independent measurements.  Whether these tensions are signaling the presence of unaccounted for systematic errors in some of the cosmological analyses, that there is new physics (i.e., beyond the standard model), or that they are merely the result of statistical fluctuations, is presently unclear and the subject of much investigation.

Although the $\Lambda$CDM model does remarkably well at describing observations of our Universe on large scales (tens to hundreds of Mpc, modulo the recent mild tensions described above), the last decade or so has seen a ramping up of detailed comparisons on smaller scales (typically kpc to Mpc), some of which have reported significant tensions with the predictions of $\Lambda$CDM-based cosmological simulations.  Three of the most widely discussed tensions are the `cusp-core' problem \citep{Flores1994, Moore1994}, the `missing satellites' problem \citep{Klypin1999, Moore1999} and the `too-big-to-fail' problem \citep{Boylan-Kolchin2011}.
The common thread between these tensions is that the $\Lambda$CDM model appears to predict too much structure (and too high densities) on small scales compared with what is inferred from observations.  These tensions are often evaluated with respect to gravity-only calculations in the context of $\Lambda$CDM, but recent work has suggested that neglecting important baryonic physics may have significant implications for these tensions.  For example, including processes such as reionisation, star formation, stellar feedback through supernovae and winds, and feedback associated with active galactic nuclei in the models have been shown to help alleviate some of these problems with $\Lambda$CDM on small scales (see e.g. \citealt{Efstathiou1992,Bullock2000,Benson2002,Mashchenko2008, Pontzen2012,Wetzel2016,Sawala2016}).

Nevertheless, it is worthwhile to consider other possibilities, which may act in conjunction with baryon physics, for these small scale problems, not least because the modelling of baryon physics on small scales is quite complex and requires a degree of fine tuning to resolve some of the aforementioned tensions.  Also, it is important that we continue to test the standard model as rigorously as possible on small scales, in order to shed light on the nature of dark matter.  

The apparent deficit of small-scale structure has led to the study and development of many extensions to the $\Lambda$CDM model which aim to reconcile these differences.  Promising extensions to the current $\Lambda$CDM paradigm, such as warm dark matter (WDM) and self-interacting dark matter (SIDM), can reduce the formation of structure on small scales.  These different cosmological models focus on changing one of the assumed aspects of dark matter.  In the case of the former, dark matter decouples in the early universe whilst still relativistic leading to non-negligible thermal velocities and free-streaming, while in the latter case dark matter is allowed to have strong self-interactions (scattering).  Both of these extensions have been shown to have success in alleviating the outlined challenges which exist with $\Lambda$CDM (see e.g. \citealt{Colin2000, Lovell2012} for the case of WDM and \citealt{Spergel2000, Zavala2013} for the case of SIDM). One further extension which has not been studied quite so extensively is a running scalar spectral index. A running spectral index is different from the previous two extensions as it does not alter the nature of dark matter but instead changes the initial conditions for structure formation, which is motivated by certain classes of inflation models. Previously this has also been demonstrated to be a promising candidate for reducing the formation of small-scale structure (see e.g. \citealt{Garrison-Kimmel2014, Stafford2020b}).

In a recent paper \citep{Stafford2020b}, we compared how effective these three different extensions were in altering various small-scale structure statistics.  To do this we used gravity-only cosmological simulations with the values adopted for the additional parameters associated with each extension being guided by current observational constraints. 
We found that all of the models can have similar effects on certain statistics,  such as the abundance of satellite galaxies inside hosts (which is one of the primary tests of $\Lambda$CDM on small scales). 
As such, it is of interest to explore new observables which could help differentiate these models and the effects they have on structure formation.
 
In this study, we exploit the fact that these different models alter the non-linear matter power spectrum along with its redshift evolution in different ways (as we will show).  Consequently, observational probes which are directly sensitive to the non-linear matter power spectrum may provide a strong test of these extensions and of $\Lambda$CDM. 
One potentially promising observable is the cosmic shear power spectrum.  As the light emitted from distant galaxies travels towards us on Earth, its path becomes distorted due to intervening matter, a phenomenon known as gravitational lensing.
One can use the correlated effect this has on galaxy shapes to extract information about the non-linear matter power spectrum. This can be done either through the two-point auto-correlation function of galaxy shapes or its Fourier analogue, the cosmic shear power spectrum. 
It is this latter statistic that we examine in this paper and how it is affected by SIDM, WDM and a running spectral index in comparison with the \LCDM result.

The impetus for this study stems from the increasing quantity and quality of cosmic shear observations being made by current Stage-III surveys, such as the Dark Energy Survey (DES)\footnote{\href{https://www.darkenergysurvey.org/}{https://www.darkenergysurvey.org/}} \citep{DES2021results},
the Hyper Suprime-Cam Subaru Strategic Survey (HSC)\footnote{\href{https://hsc.mtk.nao.ac.jp/ssp/survey/}{https://hsc.mtk.nao.ac.jp/ssp/survey/}} \citep{Hamana2020}, 
and the Kilo-Degree Survey (KiDS)\footnote{\href{http://kids.strw.leidenuniv.nl/overview.php}{http://kids.strw.leidenuniv.nl/overview.php}} \citep{Asgari2021}, and forthcoming Stage-IV cosmic shear surveys, such as \Euclid\footnote{\href{https://www.euclid-ec.org/?page_id=2581}{https://www.euclid-ec.org/?page\_id=2581}} \citep{Euclid2020},
the Rubin Observatory Legacy Survey of Space and Time (\LSST)\footnote{\href{https://www.lsst.org/scientists/}{https://www.lsst.org/scientists/}} \citep{Zhan2018}, and the Nancy Grace Roman Space Telescope (\NGRST)\footnote{\href{https://www.jpl.nasa.gov/missions/the-nancy-grace-roman-space-telescope}{https://www.jpl.nasa.gov/missions/the-nancy-grace-roman-space-telescope}} \citep{Spergel2015}.   Stage-IV surveys will greatly improve on current observations, by covering a much larger area of the sky and/or being significantly deeper, ultimately resulting in greatly improved measurements of the cosmic shear power spectrum.  

To date most of the forecasting work for Stage-IV surveys has been with regards to anticipated constraints on the standard model of cosmology and extensions that affect large-scale structure (such as evolving dark energy and massive neutrinos).  Only a small number of studies have been dedicated to the possible constraints that may be obtained for small-scale extensions.  For example, \citet{Markovic2011} explored how future cosmic shear surveys can be used to place constraints on the mass of thermal relic particles.  More recently, \citet{Hubert2021} explored constraints on decaying dark matter models.  The present paper seeks to examine the impact of the three cosmological extensions described above (WDM, SIDM, and a running scalar spectral index) on the cosmic shear power spectrum using the non-linear matter power spectrum extracted from numerical simulations.  We then explore the prospect of these upcoming Stage-IV surveys in differentiating these extensions from $\Lambda$CDM.

The paper is structured as follows.  In Section \ref{sec:extensions} we describe in more detail the extensions studied in this paper. We also discuss the numerical simulations which we use as well as how we create a non-linear matter power spectrum which covers the full dynamic range of interest for this weak lensing study. 
In Section \ref{sec:tomo_weak_lensing} we discuss how we compute the cosmic shear power spectrum and the associated uncertainties. In Section \ref{sec:results} we present our results for the cosmic shear auto- and cross-correlation power spectra for each cosmological model and finally in Section \ref{sec:conclusions} we discuss and summarise our results. 

\section{Cosmological Extensions}
\label{sec:extensions}
In this study, we analyse three separate cosmological extensions, a running scalar spectral index, warm dark matter and self-interacting dark matter. We briefly describe these in the following subsections and refer the interested reader to \cite{Stafford2020b} for further details.
Note, however, in \cite{Stafford2020b} we only examined two WDM models and two SIDM models, whereas in this study we extend our analysis slightly and look at an additional, more extreme, case for both of these extensions.
\subsection{Extensions}
\label{extensions}
\subsubsection{Running of the scalar spectral index}
In the standard model of cosmology, the power spectrum for scalar perturbations generated by inflation is assumed to follow a simple power law of the form \citep{Guth1981, Kosowsky1995}:
\begin{equation}
\label{eq:inflation}
    P(k) = Ak^{n_s},
\end{equation}
where $A$ is the amplitude of the primordial matter power spectrum, $k$ is the wavenumber and $n_s$ is the spectral index.  A first order extension is that the spectral index has some level of scale dependence, which results in a modification to the functional form of the primordial matter power spectrum \citep{Kosowsky1995}:
\begin{equation}
    \label{eq:running_pow_spec}
    P(k) = A_s(k_\textrm{pivot}) \left(\frac{k}{k_\textrm{pivot}}\right)^{n_s(k_\textrm{pivot})+\frac{\alpha_s}{2}\ln\left(\frac{k}{k_\textrm{pivot}}\right)}.
\end{equation}

Here, $\alpha_s$ is termed the `running' of the scalar spectral index and is defined as $d n_s(k)/d\ln(k)$\footnote{Although in the $\Lambda$CDM model $\alpha_s$ is assumed to be zero, virtually all models of inflation predict some level of scale dependence for $n_s$. However, the simplest single field slow-roll inflation models predict that this scale-dependence should only be of the order $10^{-3}$ \citep{Kosowsky1995}.}.
Here, $n_s(k)$ is still equal to the logarithmic slope of the power spectrum, however, it now has a $k$-dependence due to the $(\alpha_s/2)\ln(k/k_\textrm{pivot})$ term in equation \ref{eq:running_pow_spec}.
The value adopted for $k_\textrm{pivot}$ here corresponds to that for the Planck satellite mission, $k_\textrm{pivot}$ = 0.05 Mpc$^{-1}$, and corresponds to the $k-$scale at which values for $A_s$ and $n_s$ are quoted.
In this work we explore the effects that a positive value for the running (with $\alpha_s = 0.00791$) and a negative value for the running (with $\alpha_s = -0.02473$) have on the non-linear matter power spectrum and how these propagate through to an observable impact on the cosmic shear power spectrum. 
Note the cosmological parameter values chosen for the running simulations are in the context of a Planck 2015 cosmology; the maximum-likelihood values are computed from the Markov chains which includes $\alpha_s$ as a free cosmological parameter in the analysis.  This results in slight differences to the quoted values of the other cosmological parameters in Section \ref{sec:simulations} (a table of all of the parameter values for each cosmology can be found in \citealt{Stafford2020b}). The values chosen for $\alpha_s$ are discussed in detail in \cite{Stafford2020a}, but in summary represent the $\pm2\sigma$ values of the posterior distribution extracted from the Planck 2015 Markov chains\footnote{\href{http://pla.esac.esa.int/pla/\#home}{http://pla.esac.esa.int/pla/\#home}} \citep{Planck2014}. 

These two cosmological models with a running scalar spectral index were chosen in \cite{Stafford2020a}, prior to the release of the Planck 2018 cosmological parameter constraints \citep{Planck2018}.  The updated CMB constraints on $\alpha_s$ are tightened somewhat to $\alpha_s = -0.0045 \pm 0.0067$ (68\% CL TT,TE,EE+lowE+lensing) making the posterior distribution for $\alpha_s$ used in \cite{Stafford2020a} slightly out of date.
However, it is worth noting that, even with the updated CMB constraints from the Planck team, a mildly negative value for the running of the spectral index is not ruled out (see the discussion in section 7.2.1 in \citealt{Planck2018}). 
Furthermore, updated constraints from the Lyman-$\alpha$ forest find a $\approx$3$\sigma$ detection for a negative $\alpha_s$ with $\alpha_s = -0.010 \pm 0.003$ \citep{Palaqnue-Delabrouille2020}\footnote{Note, however, that \cite{Palaqnue-Delabrouille2020} marginalise over neutrino mass at the same time as $\alpha_s$, whereas the Planck analysis fixed $\Sigma M_\nu$ to 0.06 eV when constraining $\alpha_s$.}.
Therefore, the values we simulate are still compatible with current observations, even if they lie on the more extreme end of current constraints. Furthermore, as discussed later on in the paper in the context of warm dark matter models, this work seeks to answer the question whether cosmic shear can be used as a complementary probe to place \textit{independent} constraints on these additional cosmological parameters ($\alpha_s$, $M_{\textrm{WDM}}$, $\sigma/m$).

\subsubsection{Warm dark matter}

Another assumption of the standard model of cosmology is that dark matter decoupled from the primordial plasma after it became non-relativistic. This results in negligible thermal velocities at early times (hence ``cold'' dark matter) and would be expected if dark matter was composed of particles with masses in the GeV range (or larger), such as the currently favoured candidate, the WIMP (Weakly Interacting Massive Particle). 
If, however, dark matter is made up of lighter particles with masses in the keV range, such as thermal relic sterile neutrinos, the dark matter particles decouple whilst still relativistic. This type of model is referred to as a thermal relic warm dark matter (WDM) model. 
The resulting larger thermal velocities that these dark matter particles have (compared to CDM) at early times allows them to free-stream out of density perturbations. The free-streaming of dark matter particles works to suppress the growth of structure on small scales \citep{Bond1983, Bardeen1986}. 

The suppression of small-scale density perturbations due to the thermal velocity associated with the dark matter particles leads to a characteristic cut-off in the WDM power spectrum below a $k-$mode corresponding to the free-streaming scale. In this work we model the suppression of the initial linear matter power spectrum as a transfer function relative to the corresponding cold dark matter power spectrum:

\begin{equation}
    \label{eq:wdm_transfer}
    P_{\textrm{WDM}}(k) = T^2_{\textrm{WDM}}(k)P_{\textrm{CDM}}(k).
\end{equation}

We compute $T_{\textrm{WDM}}$ using the fitting formula developed in \cite{Bode2001}:
\begin{equation}
    \label{eq:bond_transfer}
    T_{\textrm{WDM}}(k) = \left[1+(\alpha k)^{2\nu}\right]^{-5/\nu},
\end{equation}
\noindent where $\nu$ is a fitting constant and $\alpha$ dictates the scale of the cut-off in the power spectrum, with this being dependent on the mass of the thermal relic particle.  We follow \cite{Viel2005}, adopting $\nu$ = 1.12 and (assuming the WDM is composed of thermal relics) and computing $\alpha$ as:
\begin{equation}
    \label{eq:alpha}
    \alpha = 0.049 \left(\frac{M_{\textrm{WDM}}}{1\textrm{keV}}\right)^{-1.11} \left(\frac{\Omega_{\textrm{WDM}}}{0.25}\right)^{0.11}
    \left(\frac{h}{0.7}\right)^{1.22}\textrm{Mpc}\:h^{-1},
\end{equation}
\noindent where $M_{\textrm{WDM}}$ corresponds to the mass of the WDM particle, $\Omega_{\textrm{WDM}}$ is the present-day density of WDM
\footnote{We assume all of the dark matter is in the form of WDM and so $\Omega_{\textrm{WDM}} = \Omega_{\textrm{CDM}}$.}
in units of the critical density and $h$ is the reduced Hubble's constant.

It is evident from equation \ref{eq:alpha} that for lighter WDM particles $\alpha$ increases, pushing the scale of the cut-off in the matter power spectrum to smaller $k-$modes (larger physical scales). In this study we examine $M_{\textrm{WDM}}$ = (0.5, 2.5, 5.0) keV. Our choice for the parameter values investigated in this study aims to bracket the current observational constraints placed on the mass of a WDM particle from different probes such as the Lyman-$\alpha$ forest \citep{Viel2013, Irsic2017}, the Milky-Way's satellite population \citep{Lovell2014, Kennedy2014, Jethwa2018, Nadler2019, Nadler2020} and time-delay measurements of strongly gravitationally lensed quasars \citep{Hsueh2019, Gilman2019}. Although the 0.5 keV model is perhaps currently in tension with constraints from the previous observations, we examine it here to see if cosmic shear, as an independent test with very different systematics compared to previous methods, can also rule out such a model. 

\subsubsection{Self-interacting dark matter}

In terms of the possible interactions that dark matter particles can experience, the standard model of cosmology adopts the simplest assumption that dark matter interacts only via gravity and is therefore `collisionless'.   The final extension that we investigate is a relaxation of this assumption, allowing the dark matter particles to have strong self-interactions.
SIDM was proposed to alleviate the `cusp-core problem' \citep{Flores1994} by \cite{Spergel2000} and has indeed been shown in the literature to produce strong cores inside dark matter haloes and can have particularly large effects on satellite subhaloes, strongly reducing their masses (see e.g. \citealt{Penarrubia2010, Vogelsberger2012, Dooley2016}).  We explore how these effects translate through to the non-linear matter power spectrum.

To study the effect of strong self-interactions we use a version of the \texttt{GADGET3} N-body code (discussed in Section \ref{sec:simulations}) which was modified in \cite{Robertson2019} to include dark matter self-interactions, which are assumed to be elastic. 
The additional parameter this adds to the standard 6 free parameters of the $\Lambda$CDM model is ($\sigma/m$), which is the cross-section for interaction of dark matter particles. In this study the values which we explore are ($\sigma/m$) = (0.1, 1.0, 10.0) cm$^{2}$ g$^{-1}$.  Again, our choices for the values adopted for this cosmological parameter are guided by the current observational constraints. In the case of the self-interaction cross-section, these constraints are placed by probes such as perturbations in strong lensing arcs \citep{Meneghetti2001, Robertson2019}, dark matter-galaxy off-sets in colliding galaxy clusters \citep{Randall2008, Kahlhoefer2015, Harvey2015, Kim2017, Robertson2017, Wittman2018}, cluster shapes \citep{Miralda-Escude2002, Peter2013} group shapes \citep{Sagunski2021} and also from subhalo evaporation arguments \citep{Gnedin2001}. Furthermore, \cite{Banerjee2020} showed that constraints of the order $\sigma/m$ $\la$ 2 cm$^2$ g$^{-1}$ can be placed on the cross-section for self interaction when combining observations of the distribution of subhaloes and weak lensing measured density profiles.

The largest cross-section we examine is somewhat in tension with some of the observations listed here, but it is worth noting that these are primarily focused on large scales (galaxy clusters). There is evidence to suggest that SIDM models with cross-sections as high as 50 cm$^{2}$ g$^{-1}$ are viable when using constraints coming from dwarf galaxies \citep{Elbert2015, Correa2021}.  As our focus is on small scales, the range of cross-sections we explore is therefore plausible given current constraints.
Note, although the simulation code developed by \cite{Robertson2019} has the functionality for scattering events to be both angular and velocity dependent, for simplicity we focus on the case where scattering events are velocity-independent and isotropic.  

\subsection{Simulations}
\label{sec:simulations}

The simulations used in this study are those first introduced in \cite{Stafford2020b}.  All of the simulations follow dissipationless physics only, meaning they follow either just the gravitational evolution (for CDM, WDM, and the running spectral index models) or the gravitational and self-scattering (SIDM) evolution.  The impact of physical processes associated with the baryonic component (i.e., radiative cooling, star formation, feedback processes) is not included but is discussed further in Section \ref{sec:conclusions}.

The simulations are run with the \texttt{GADGET3} code (last described in \citealt{Springel2005b}).  Each simulation is 25 comoving \Mpcph~on a side and contains 1024$^3$ dark matter particles. 
We adopt a fixed physical gravitational softening length of 250 $h^{-1}$ pc at $z$ $\leq$ 3, with this being a fixed comoving length at higher redshifts.
The particle mass for all of the simulations, with the exception of the two with a running spectral index, is $m_{\textrm{DM}}$ = 1.266 $\times$ 10$^6$ M$_{\odot}$ $h^{-1}$.  For the two `running' simulations, the particle mass is slightly different, owing to the slightly different values adopted for $\Omega_\textrm{m}$ and $h$ (see table 1 in \citealt{Stafford2020b}).
Alongside these main simulations, we also use a secondary suite of simulations of varying box size and resolution. These include a suite of simulations which are 400 comoving \Mpcph~on a side, as well as a suite which are 100 comoving \Mpcph~on a side. Both suites contain 1024$^3$ collisionless particles. Note, however, that we only run the cosmologies with a running scalar spectral index (as well as the reference \LCDM cosmology) in these larger volumes, as they are required for the process of creating a spliced power spectrum (see Section \ref{sec:constructing_mat_pow}).

The initial conditions (ICs) for the simulations are generated using a modified version of the \texttt{N-GenIC}\footnote{\href{https://github.com/sbird/S-GenIC}{https://github.com/sbird/S-GenIC}} code \citep{Springel2005a} which was modified to include second-order Lagrangian Perturbation Theory corrections. The ICs are generated at a starting redshift of $z$ = 127 with each simulation being initialised with the same random phases. 
The input linear theory matter power spectrum and transfer functions are computed with the Boltzmann code \texttt{CAMB} (\citealt{Lewis2000}, August 2018 version). 
The initial conditions and the background expansion rate in the simulations are computed assuming the Planck 2015 maximum-likelihood cosmological parameters \citep{Planck2014} ($H_0 = 67.31$ km s$^{-1}$ Mpc$^{-1}$; $\Omega_{\textrm{DM}}=0.264$; $\Omega_\textrm{b}=0.049$; $n_s=0.966$; $\sigma_8 = 0.830$; $\Sigma M_{\nu}=0.06$ eV), with the exception of the two running simulations (as explained above). 

Note that we use a version of the \texttt{GADGET3} code (see \citealt{McCarthy2018}) that includes the impact of massive neutrinos on the expansion rate and the growth of fluctuations (i.e., accounts for their free streaming), using the semi-linear algorithm of \citet{Ali-Ha2013}.  As the simulations employed here adopt the minimum allowed neutrino mass ($\Sigma M_{\nu}=0.06$ eV), consistent with what is assumed in the Planck analysis, their incorporation will not have important consequences for the present study, but we include them for consistency.

\subsection{Matter power spectra}

In this study we probe the theoretical cosmic shear power spectrum over a large range of multipoles/angular scales.  This requires that we have a model for the non-linear matter power spectrum which spans a large range of wavenumbers/physical scales.  However, the high-resolution 25 \Mpcph~ simulations that we use to probe the non-linear effects only have a $k$-range spanning from $\approx$ 0.25 \hpMpc~ to 257.36 \hpMpc.  While the simulations extend to large enough $k$ (i.e., small scales) for upcoming lensing surveys, they are clearly too small to capture all of the relevant structure on large scales.  Even though our interest is primarily focused on small scales, we nevertheless want to construct realistic synthetic cosmic shear observations.  (Furthermore, as we will show, the running of scalar spectral index models have important contributions from large scales.)  We thus require a way of modelling the non-linear matter power spectrum over a wide range of scales.

\subsubsection{Method for constructing matter power spectra}
\label{sec:constructing_mat_pow}

In order to model the non-linear matter power spectrum over a wide range of wavenumbers, we combine our direct simulation measurements with predictions from \texttt{Halofit} \citep{Smith2003, Takahashi2012} on larger scales.  Briefly, the \texttt{Halofit} algorithm provides an empirical correction for scaling the linear power spectrum (e.g., from a Boltzmann code) for a given cosmology to the non-linear power spectrum.  The empirical corrections were derived from fits to a large suite of CDM-based cosmological (N-body) simulations.  We use the Boltzmann code \texttt{CAMB} to compute the \texttt{Halofit} prediction at a given redshift.  This provides us with the fiducial non-linear $\Lambda$CDM power spectrum.  To construct a full power spectrum for each of our extensions to $\Lambda$CDM, we use the ratios of the non-linear matter power spectra extracted directly from the simulations (with respect to our $\Lambda$CDM case) as a multiplicative `boost factor' for the non-linear \texttt{Halofit} \LCDM prediction via:
\begin{equation}
\label{eq:boost_factor}
    P_{C}(k, z) = P^{\texttt{Halofit}}_{\Lambda \rm{CDM}}(k, z) R_{C}(k, z),
\end{equation}
where $P_{C}(k, z)$ is the non-linear matter power spectrum in a given cosmological extension, and $R_{C}(k, z)$ is the constructed ratio (boost factor) for a given cosmology.

This approach allows us to combine multiple simulations of varying box size or resolution (by combining their ratios), following a similar vein to the power spectrum splicing in previous Lyman-alpha forest work (e.g.,  \citealt{Palanque-Delabrouille2015}).  The combined ratios can then be used to seamlessly scale the power spectra derived on very large scales (e.g., via linear theory, perturbation theory, \texttt{Halofit}, etc.) via equation~\ref{eq:boost_factor} to produce an absolute power spectrum for a cosmological model.

Note that previous studies have shown that cosmological simulations which do not simulate the smaller $k-$modes (larger physical scales) do not accurately represent the intermediate $k-$modes sampled in the cosmological volume (see e.g. \citealt{Power2006, Heitmann2010, Klypin2018, Knabenhans2019}), making it difficult to combine the absolute matter power spectra from simulations of different box sizes.  This motivates the use of combining ratios of power spectra rather than splicing the absolute power spectra themselves.  We have explicitly tested how sensitive the ratio of the different power spectra are to resolution as well as box size effects (see Fig.~\ref{fig:res_test} in Appendix \ref{sec:resolution_test}), concluding that they are more robust to such effects than are the absolute power spectra.  

To explain how we construct the non-linear matter power spectrum in a bit more detail, firstly, we compute the matter power spectra for the different simulations using \texttt{GENPK}\footnote{\href{https://github.com/sbird/GenPK}{https://github.com/sbird/GenPK}} \citep{Bird2017}.
We then re-bin the matter power spectra to have 10 $k$-modes per bin (corresponding roughly to rebinning the power spectra in logarithmic bins of width 0.0143 dex), in order to smooth out some of the associated noise.  
We compute the mean power in each bin, as well as the mean wavenumber.  The ratios of the power spectrum for each cosmological extension with respect to the \LCDM result are then computed. 
In the case of the SIDM and WDM models, all six of these simulations have ratios which tend to 1 within the simulated volume (i.e., the physical effects of these modifications are confined to small scales), as such we simply extrapolate these ratios to smaller $k$-values (larger physical scales) assuming they are fixed at unity over the entire range. 
In the case of the two simulations which have a running scalar spectral index, the ratios of these power spectra do not tend to unity within the high-resolution 25 \Mpcph~boxes, as such we need to simulate the ratio over a wider $k$-range. 
To do this we use two further sets of simulations of size 100 \Mpcph~and 400 \Mpcph, each with 1024$^3$ dark matter particles, to compute the ratio of these models with respect to a corresponding \LCDM result. 
As a result, we can accurately probe the non-linear matter power spectrum in these simulations up to approximately linear scales where we can then make use of the theoretical prediction from \texttt{Halofit}. 
The reason we do not use the \texttt{Halofit} prediction over the entire $k$-range for the simulations with running is because it does not tend to reproduce the effects we see in the running simulations on non-linear scales (as shown by the dotted line in Fig.~\ref{fig:res_test}), discussed in \cite{Stafford2020a} (see also \citealt{Smith2019}).

With a combined ratio spanning the entire desired $k$-range for each cosmological model, we fit a cubic spline to the the data, which is smoothed with a 3rd order Savitsky-Golay filter \citep{Savitzky1964} over the nearest 51 wavenumbers.  These two steps are done to ensure a smooth continuous function describing the ratio over the entire $k$-range. 
We repeat this process at each redshift for which we have a simulation snapshot.
Once we have the ratio for each cosmology at each redshift relative to the \LCDM prediction, we use it as a multiplicative boost factor to a \LCDM matter power spectrum (at the corresponding redshift) to obtain the absolute power spectrum for each cosmology.  Note that we have simulation snapshots at $z = \{0.0, 0.125, 0.25, 0.375, 0.5, 0.75, 1.0\}$ for which we compute the non-linear total matter power spectrum. 
When we compute our shear power spectra described in Section \ref{sec:theory}, we use a cubic spline to interpolate between both redshift and $k-$modes. We set the power to zero for $k-$modes outside of the range sampled by our constructed matter power spectrum.

\begin{figure}
    \centering
    \includegraphics[width=\columnwidth]{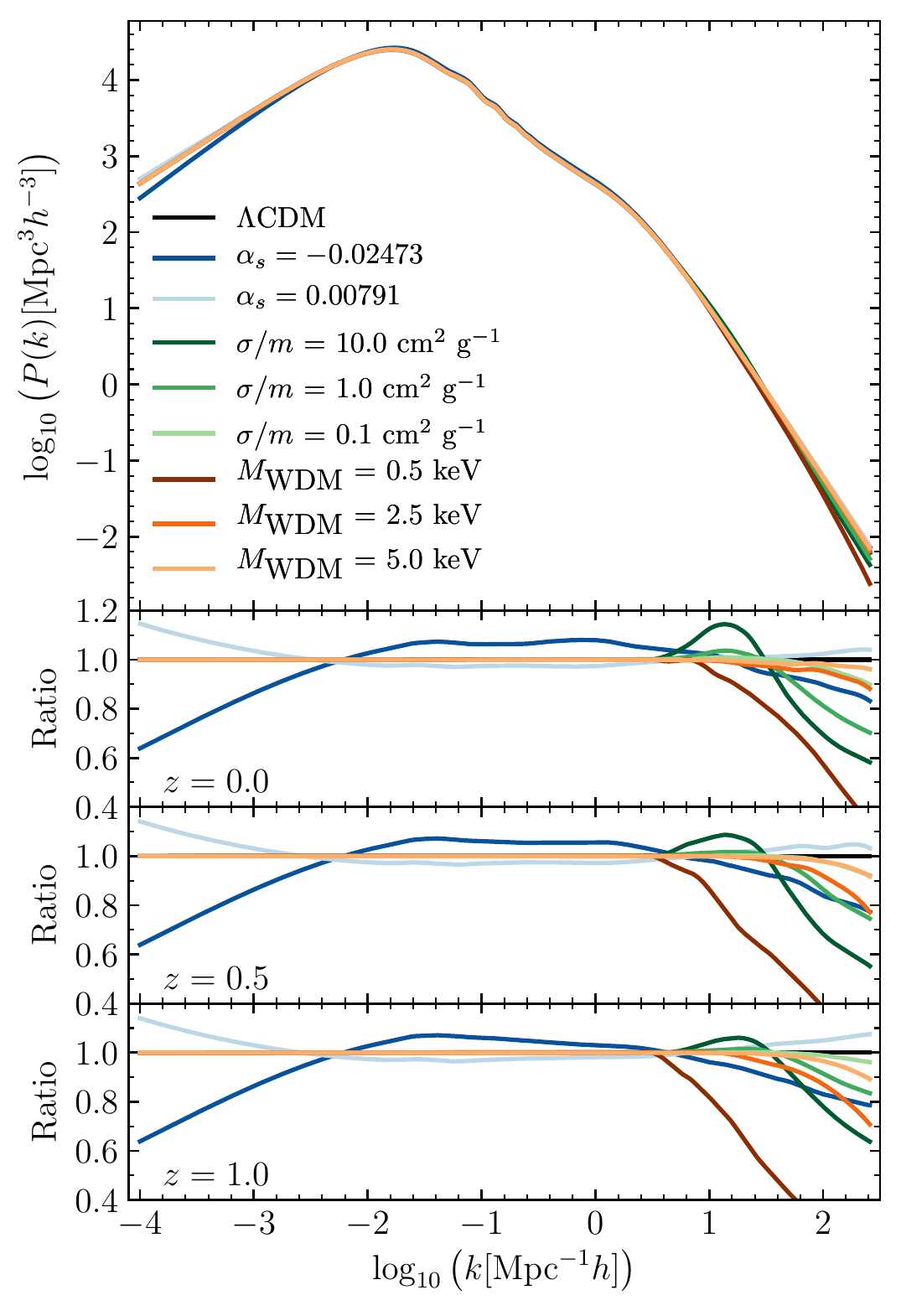}
    \vspace{-0.5cm}
    \caption{Top: the constructed non-linear matter power spectrum for each cosmological extension indicated with varying colours, computed at $z$ = 0. Bottom (three panels): the matter power spectrum of each cosmological extension normalised to the \LCDM~case. Redshifts 0, 0.5 and 1.0, respectively, are shown, demonstrating the redshift evolution of the ratios for each cosmological model. All models, with the exception of a cosmology with a positively running scalar spectral index, result in some level of suppression of the non-linear matter power spectrum compared to the \LCDM~prediction on small physical scales. This suppression ranges from around 5\% in the case of the less extreme WDM cosmology to $\approx$ 60\% for the more extreme WDM cosmology for the smallest scales examined here at the present day.}
    \label{fig:non_linear_Pk}
\end{figure}

\subsubsection{Resulting non-linear matter power spectra}
\label{sec:mat_pow}

The resultant constructed matter power spectra can be seen in Fig.~\ref{fig:non_linear_Pk}, where we show the absolute power spectra at redshift $z$ = 0 in the top panel, and the redshift evolution of the ratios computed with respect to the \LCDM result in the bottom three panels. 
We only show here the results up to a maximum redshift of $z = $ 1.0 as this is the maximum source redshift we use in our weak lensing calculations (we discuss our choice of a maximum source redshift of $z = $ 1.0 in Section \ref{sec:l_max_4000}).

A feature common to most of the cosmological extensions examined here is the suppression of small-scale power at $k > 10$ \hpMpc.  A cosmological model with a negatively running scalar spectral index has less power on small scales due to a dampening effect on the initial density perturbations generated by inflation.
WDM cosmologies have small-scale density perturbations erased due to free streaming.
Finally, SIDM cosmologies have small-scale clustering erased at late times due to the self-interactions creating cores (and generally erasing structure) in otherwise high-density dark matter haloes.
In detail, however, differences between the three models do exist. For example, whereas the suppression in the matter power spectrum decreases with decreasing redshift for WDM and a negative value for $\alpha_s$, the suppression increases with time in a SIDM cosmology. 
The suppression reduces with time for the other two cosmological models due to mode mixing transferring power from large scales to smaller scales.
A similar result to this was seen in \cite{Stafford2020a} for the case of a running spectral index and in \cite{Viel2012} for the case of WDM.
However, in the case of SIDM the effect seen in the non-linear matter power spectrum of the suppression becoming larger with decreasing redshift, which is a feature unique to SIDM, occurs due to the higher virial velocities present at late times. 
This causes scattering events to be more efficient at redistributing the mass inside the inner regions of haloes. Furthermore, there is a cumulative number of scattering events which occur with decreasing redshift.
These combined effects result in the suppression in the matter power spectrum extending to smaller $k$-modes.
The differences seen in the redshift evolution of the matter power spectrum are important as measurements at a single fixed redshift could result in different cosmological models having very similar effects. 
See, for example, the $z$ = 0.5 panel in Fig.~\ref{fig:non_linear_Pk} where the $M_{\textrm{WDM}}$ = 5.0 keV and the $\sigma/m$ = 0.1 cm$^2$ g$^{-1}$ are predicted to have an almost indistinguishable effect on the non-linear matter power spectrum. 
However, at earlier and later redshifts the predictions for their suppression of the matter power spectrum are noticeably different. 

The model with a positive running scalar spectral index is the only model to predict an enhancement in power on these same scales, except for the slight increase in power seen in the more extreme SIDM models on $k-$scales $\approx$ 25 \hpMpc. The increase in power on small-scales in the SIDM models is due to the re-distribution of dark matter in the central regions of dark matter haloes.  In particular, the enhancement mainly occurs due to the outward scattering of dark matter particles from the very central regions to somewhat larger radii (see figure 5 in \citealt{Stafford2020b}, for example). This outer radii, where particles tend to gather in stable orbits corresponds approximately to the radius where we expect each particle to have interacted at least once per Hubble time (see section 5.3 of \citealt{Rocha2013}, for example, for a discussion on this).

One obvious feature of the two cosmologies with a running scalar spectral index is the cross-over regions seen in the ratio panels (in the range $-2 \leq \log_{10}(k [h\textrm{Mpc}^{-1}]) \leq 1$).
The reason these cross-over regions exist is because of differences in the amplitude of the primordial matter power spectrum, $A_s$, which is larger in the case of the positive running cosmology and smaller in the negative running case with respect to $A_s$ for the fiducial $\Lambda$CDM case.  (Note that $A_s$ varies between the models in order to retain a good match to the Planck CMB angular power spectrum, see discussion in \citealt{Stafford2020a}.)  This amplitude difference produces an additional offset with respect to the \LCDM power spectrum on top of the the resulting increase (decrease) in power on large and small scales due to the positive (negative) running of the spectral index.

\section{Tomographic Weak Lensing}
\label{sec:tomo_weak_lensing}

In this section we describe our methodology for computing cosmic shear power spectra from the non-linear matter power spectra described above and how we generate noisy realisations (synthetic observations) of the cosmic shear power spectra.

\subsection{Theory}
\label{sec:theory}

Weak lensing describes the deflection of light rays due to the presence of large-scale structure in the Universe, which results in slight correlated distortions in the observed shapes of galaxies \citep{Blandford1991, Miralda-Escude1991, Kaiser1992}.
The lensing of galaxies leads to two effects: the dilation or magnification of an image which can be described by the convergence, $\kappa$, and the stretching (shearing) of an image, $\gamma_{1,2}$.  In this study we focus on the shears of galaxies which can be used to probe the projected mass distribution via the galaxy shape correlation functions.  Additional information on the growth of structure over cosmic time can be obtained if one has redshift measurements of the background galaxies. 
In this case, the source distribution can be discretized into redshift (or tomographic) bins, allowing one to probe the three-dimensional matter distribution (see \citealt{Kilbinger2015} for a recent review). 

The 2-point correlation function of galaxy shapes, and its Fourier analogue the power spectrum, are directly linked to the underlying matter distribution and its power spectrum integrated along the line of sight. 
We compute the cosmic shear power spectrum via: 
\begin{equation}
\label{eq:shear_pow}
P^{\gamma}_{ij}(\ell) =  \int^{\chi_{\rm{H}}}_{0}d\chi_{\rm{l}}(1+z_{\rm{l}})^2 W_i(\chi_{\rm{l}})W_j(\chi_{\rm{l}})P_{\rm 3D} \left(k=\frac{\ell}{\chi_{\rm{l}}}, \chi_{\rm{l}}\right) \ \ \ ,
\end{equation}

\noindent where $\chi_{\rm{l}}$ is the comoving distance to the lens at redshift $z_{\rm{l}}$, $P_{\rm 3D}$ is the non-linear matter power spectrum, and $W_i$ and $W_j$ are the lensing efficiencies in the tomographic bins $i$ and $j$, defined as:

\begin{equation}
\label{eq:lensing_kernel}
    W_{i, j}(z_l) = \frac{3}{2}\Omega_{\rm{m}}\left(\frac{H_{0}}{c}\right)^2 \int^{z_{\rm{max}}}_{z_{\rm{l}}}\frac{\chi_{\rm{l}} - \chi_{\rm{s}}}{\chi_{\rm{s}}} n_{i, j}(z_{\rm{s}}) dz_{\rm{s}}.
\end{equation}

Here $\chi_{\rm{s}}$ corresponds to the comoving distance to the source at redshift $z_{\rm{s}}$ and $n_{i}$ is the normalized homogeneous source distribution in tomographic bin $i$ given by:
\begin{equation}
\label{eq:n_z_norm}
    n_i(z) = \frac{n(z)}{\int^{z_{\rm{max}}}_{0}n(z) dz},
\end{equation}

It is worth mentioning that equation \ref{eq:shear_pow} is predicated upon some important assumptions. 
These include the Limber approximation \citep{Limber1953, Kaiser1992}, which only includes modes in the plane of the sky, neglecting those between structures at different epochs in the line of sight integration. 
The small-angle and flat-sky approximations have also been adopted, which allows one to replace a spherical harmonics transform with a Fourier transform \citep{Hu1999}.
Other assumptions which are embedded inside the lensing efficiency $W_{i, j}$ (see equation \ref{eq:lensing_kernel}) is that of a homogeneous galaxy distribution, which ignores source-source clustering \citep{Schneider2002} as well as source-lens clustering \citep{Bernardeau1998, Hamana2002}.  Spatial flatness has also been assumed in equation \ref{eq:shear_pow}.

Although the previous assumptions may introduce some level of systematic error in the calculation of the shear power spectrum, we neglect them here owing to the fact that, firstly, they will be common to each cosmology and should therefore be less important when we focus on the ratios of the cosmic shear power spectra, and, secondly, they mainly affect the largest angular scales ($\ell$ < 10) \citep[see for example][]{Schmidt2008, Giannantonio2012, Kilbinger2017} which is not our focus.

The uncertainty in the shear power spectrum can be expressed as (\citealt{Kaiser1998,Hu1999,Euclid2020}):
\begin{equation}
\label{eq:fiducal_err}
\Delta P_{ij}^{\gamma}(\ell) = \sqrt{\frac{2}{\left(2\ell+1\right)\Delta\ell f_{\textrm{sky}}}}
\left[P_{ij}^{\gamma}(\ell) + \delta_{ij} \frac{\left<\gamma_{\textrm{int}}^2\right>}{\overline{n_i}}
\right],
\end{equation}
where $\Delta\ell$ is the multipole bandwidth, $f_{\textrm{sky}}$ is the fraction of sky surveyed, $\delta_{ij}$ is the Kronecker delta symbol, $\left<\gamma_{\textrm{int}}^2\right>$ is the shape noise, representing the variance of observed galaxy ellipticities (which we take to have a value of 0.261, motivated by \citealt{DES2021}), and $\overline{n_i}$ is the surface density of source galaxies in the tomographic bin, expressed in steradians$^{-1}$.
The term under the square root accounts for the limited number of available independent $\ell$ modes.  The first term in the square brackets corresponds to the cosmic variance and the second term is a Poisson noise term.  In addition to introducing scatter in the cosmic shear power spectrum (via the Poisson noise term), the shape noise also contributes to an additive shot noise term that biases the auto power spectrum but not the cross-power spectrum, as the shot noise in different tomographic bins is uncorrelated.  Thus, the shot noise must be subtracted from the estimated auto power spectra.

As written in equation~\ref{eq:fiducal_err} (see also \citealt{Euclid2020}, eqns.~118 and 125), it appears that the Poisson noise only applies to the uncertainty in the auto-correlation power spectrum and not the uncertainty in the cross-correlation power spectrum, in analogy to the way shot noise contributes to the measured power spectrum but not the cross-spectrum.  While the Poisson noise is uncorrelated between tomographic bins and therefore does not bias the cross-spectra, we find that there is a significant contribution to the \textit{uncertainty} in the measured cross-spectra due to Poisson noise.
Furthermore, equation~\ref{eq:fiducal_err} only provides an estimate of the Gaussian errors; i.e., the diagonal elements of the covariance matrix, but in principle there could be significant non-Gaussian contributions.  For these reasons, instead of using the standard analytic error estimate in equation~\ref{eq:fiducal_err}, we instead generate synthetic weak lensing observations using the \FLASK~software package, as described below.

We note that cosmic variance, shape noise, and Poisson errors are not the only sources of uncertainty for the cosmic shear power spectrum.  One of the major astrophysical sources of uncertainty stems from the intrinsic alignment of galaxy shapes. This stems from tidal interactions during the formation period of nearby galaxies which induces an intrinsically correlated orientation of the galaxies' shapes \citep{Joachimi2015, Kiessling2015, Kirk2015} which works to dilute the cosmological signal in the two-point correlation function of these shapes. Additionally, another important source of error stems from photometric redshifts being used for source galaxies. This leads to some galaxies being ascribed the wrong redshift and blurring the edges of tomographic bins. A comprehensive discussion on the systematic errors affecting weak lensing surveys can also be found in \cite{Mandelbaum2018}.  For simplicity we ignore these effects in the present study.

\subsection{Synthetic cosmic shear observations}
\label{sec:synth_cosmic_shear_obvs}

For our cosmic shear analysis there are several choices one needs to make to specify the setup.  For example, as seen in equation \ref{eq:shear_pow}, the shear power spectrum will directly depend on the redshift distribution of the source galaxies. It will also depend on how many tomographic bins are used in the analysis.  Furthermore, the associated uncertainty on the shear power spectrum will depend on the density of source galaxies as well as on the survey sky coverage. 

In this study we are interested in whether or not future Stage-IV weak lensing surveys such as \Euclid, \LSST, and \NGRST~would be able to rule out, or help place constraints on, the different cosmological extensions examined in this study. 
For our fiducial results we compute the shear auto- and cross-power spectra for a \Euclid-like setup, however, the survey parameters of an \LSST-like setup would be very similar, and so the results we present below would not differ much (we have explicitly verified this).  We discuss the differences with respect to an \NGRST~setup in Section \ref{sec:l_max_4000}.

When constructing the \Euclid-like setup, we follow as closely as possible that described in \cite{Euclid2020}. In particular, the source galaxy distribution is defined via:
\begin{equation}
\label{eq:n_z}
    n(z) = z^{\alpha}\exp\left(-\left[\frac{z}{z_0}\right]^{\beta}\right),
\end{equation}
\noindent where $\alpha$, $\beta$ and $z_0$ are survey specific parameters which describe the source distribution.  We adopt values of 2, 1.5 and 0.636, respectively, to be similar to the source distribution expected for the \Euclid~survey. 

The planned analysis of the \Euclid~survey splits the galaxy distribution up into 10 tomographic bins with a redshift range of $0.001 \leq z_s \leq 2.5$. 
However, in this study we focus on redshifts $\leq$ 1, which approximately corresponds to the first 6 tomographic bins in the \Euclid~setup. 
As such, we split our galaxy distribution up into 6 tomographic bins between $0.001 \leq z_s \leq 1.0$. The edges of each tomographic bin are $z_i$ = \{0.001, 0.414, 0.554, 0.669, 0.777, 0.885, 1.0\} and are defined such that there is an equal number of galaxies in each bin, essentially fixing the associated shot noise in each bin.  We choose a maximum source redshift of 1 as, for a fixed range of angular scales, higher redshift measurements correspond to larger physical scales.\footnote{Note, the minimum source redshift of 0.001 was chosen to coincide with the minimum redshift used in \cite{Euclid2020}.}  As the extensions we explore mostly affect small physical scales, we do not expect them to be easily distinguishable from $\Lambda$CDM at high redshifts (the exception to this are the models which have a running of the scalar spectral index, as we discuss below).  Specifically, we compute the cosmic shear power spectrum up to an $\ell_{\textrm{max}} = 4,000$, which lies close to the `optimistic' case for the \Euclid~survey of $\ell_{\textrm{max}}=5000$ (with their `pessimistic' case being $\ell_{\textrm{max}}$=1500, if one does not include non-Gaussian contributions to the covariance matrix). This $\ell_{\textrm{max}}$ corresponds to angular scales of around 0.44 arcmin, or $k$ [\hpMpc] = 2.18, at $z = 1$. 

Note that, as surveys such as \Euclid~and \LSST~will make use of photometric redshifts (see e.g. \citealt{Euclid2020_photo}), one needs to account for the associated uncertainty by convolving the number density distribution in equation \ref{eq:n_z_norm} with a probability distribution function accounting for uncertainties in the photometric redshifts, along with accounting for catastrophic outliers (see for example equation 115 in \citealt{Euclid2020}). These are galaxies which have a severely incorrect measurement of their redshift and have therefore been placed in the wrong tomographic bin. We ignore this effect in this paper, however.

To generate the synthetic weak lensing observations using the cosmic shear power spectra computed in Section \ref{sec:theory} and the details of the \Euclid~survey choices above, we use the publicly available software package \texttt{FLASK}\footnote{\href{https://github.com/hsxavier/flask}{https://github.com/hsxavier/flask}} \citep{Xavier2016}. \FLASK~can generate lognormal (or Gaussian) realisations of correlated fields on spherical shells.  We make use of its ability to generate weak lensing shear fields. Each field can be generated tomographically, with the statistical properties of these fields (including their cross-correlations) being defined via input angular power spectra (where one provides as input the auto- and cross-spectra that they want maps created for). We provide as input the theoretical cosmic shear power spectra as computed in equation \ref{eq:shear_pow}, which in the flat-sky approximation is equivalent to the convergence power spectrum \citep{Hu2000, Kilbinger2017, Bartelmann2001}.  We use \FLASK~to produce many realisations of our theoretical shear power spectra, allowing us to evaluate the full covariance matrix in the error analysis, rather than just using the Gaussian errors computed using equation \ref{eq:fiducal_err}, and to evaluate the potential Poisson noise contribution to cross-power spectra between tomographic bins. 
Note that the measured power spectra computed from the \FLASK~maps will contain a shot noise contribution. Therefore, when computing the full covariance matrix we subtract this shot noise from the auto-correlation power spectra.

\begin{table}
\caption{The shift parameters used to define the lognormal realisations of each shear field. The columns from left to right are the tomographic bin number, the mean redshift of the tomographic bin, the calculated shift parameter for that bin.}
\begin{center}
\begin{tabular}{|l|l|l|}
\hline
\multicolumn{1}{|c|}{$i$} & \multicolumn{1}{c|}{$z_{\textrm{mean}}$} & \multicolumn{1}{c|}{$\lambda$} \\ \hline
1                         & 0.3003                                   & 0.0048                         \\
2                         & 0.4874                                   & 0.0099                         \\
3                         & 0.6126                                   & 0.0141                         \\
4                         & 0.7233                                   & 0.0181                         \\
5                         & 0.8307                                   & 0.0224                         \\
6                         & 0.9417                                   & 0.0272                         \\ \hline
\end{tabular}
\label{Tab:redshifts}
\end{center}
\end{table}

In order to define the log-normality of the shear field, one needs to specify a shift parameter. We compute this shift parameter ($\lambda$) following \citet{Hilbert2011} who used the Millennium Simulation \citep{Springel2005b} to produce synthetic convergence and shear maps.
From these maps they measure the convergence distribution and find that it is best fit with a zero-mean shifted log-normal distribution. They do this at multiple redshifts and provide an empirical formula, which we use here, that captures the redshift evolution of the shift parameter $\lambda$ ($\kappa_0$ in \citealt{Hilbert2011}) 
\begin{equation}
\label{eq:hilbert38}
\lambda(z) = 0.008z + 0.029z^2 - 0.0079z^3 + 0.00065z^4.
\end{equation}
We substitute in the mean redshift of each tomographic bin to calculate the associated shift parameter for that field.  The values calculated can be found in Table \ref{Tab:redshifts}.  

\begin{figure*}
    \centering
    \includegraphics[width=\textwidth]{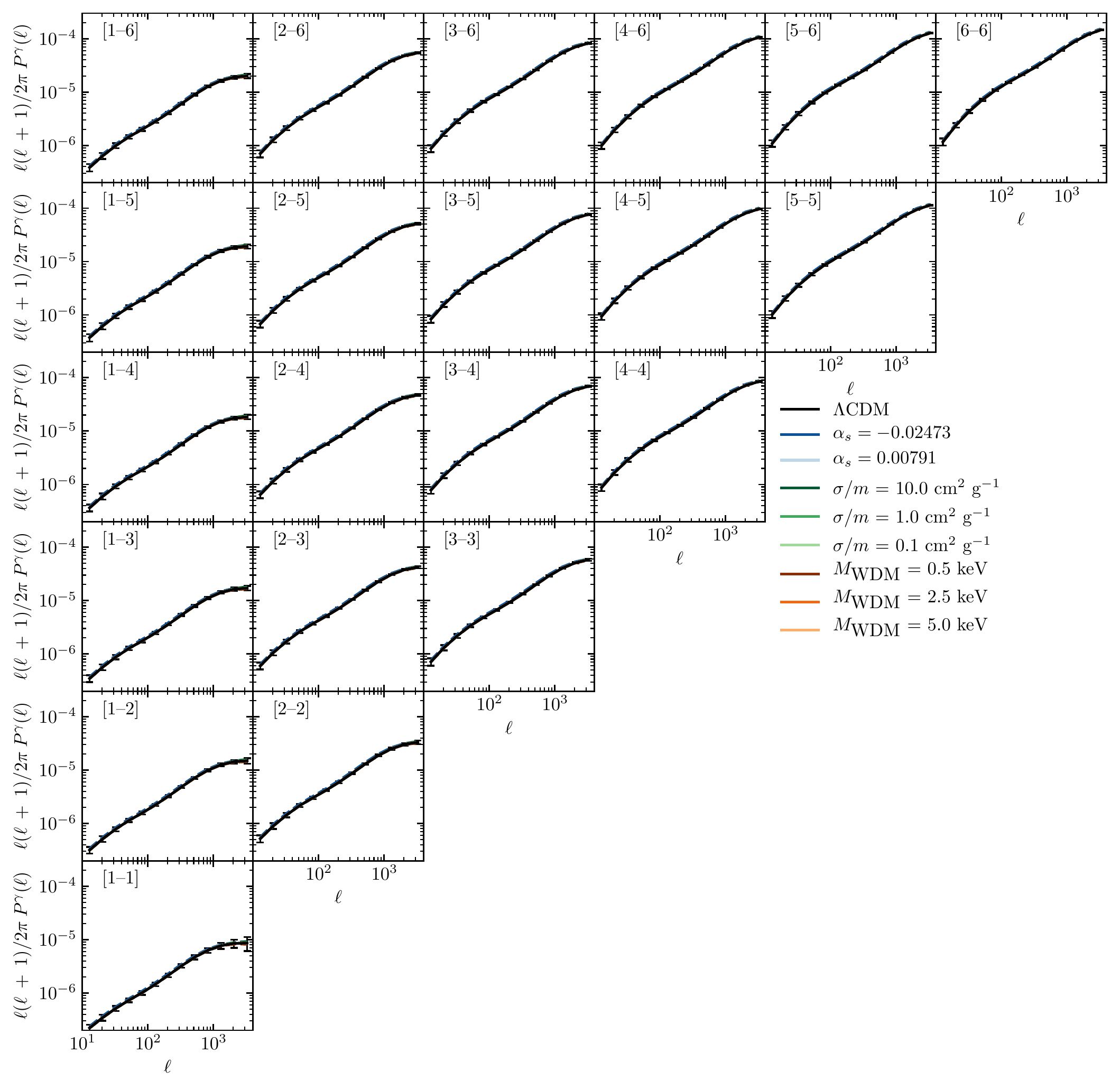}
    \vspace{-0.5cm}
    \caption{The shear auto- and cross-correlation power spectra computed for each cosmological model. The source galaxy tomographic bins are distributed between a minimum source redshift of 0.001 and a maximum source redshift of 1.0 and such that each tomographic bin has the same number density of source galaxies. We split the distribution up into 6 tomographic bins to mimic a \Euclid-like survey and plot up to $\ell_{\textrm{max}}$=4,000, which lies between the pessimistic and optimistic case for the \Euclid~survey. Here the error bars show the standard deviation of the realisations of each power spectrum evaluated using the \FLASK~package (see text).  In general, the deviations to $\Lambda$CDM case are subtle and require a quantitative evaluation of the signal-to-noise ratio.}
    \label{fig:shear}
\end{figure*}

We want \FLASK~to output shear maps which will also capture the noise associated with measurements of the cosmic shear power spectrum (see Section \ref{sec:theory}). Therefore, we also supply \FLASK~with an angular selection function, such that the map will be masked to reproduce a survey's specific $f_{\textrm{sky}}$, as well as providing a redshift selection function so that the number of galaxies in a tomographic bin is reproduced. One also needs to supply the software with a value for the ellipticity dispersion of galaxy shapes, $\left<\gamma_{\textrm{int}}^2\right>$, in order to incorporate shape noise into the shear maps created.  As already noted, we adopt $\left<\gamma_{\textrm{int}}^2\right> = 0.261$.

The maps output by \FLASK, which are output in HEALPix\footnote{\href{https://healpix.sourceforge.io/}{https://healpix.sourceforge.io/}} format, contain the mean source ellipticity in each pixel, calculated via:
\begin{equation}
\label{eq:ellip}
\epsilon(j) = g(j) + \frac{\epsilon_{s}}{\sqrt{N_{\textrm{gal}}}},
\end{equation}
\noindent where $g(j)$ is the reduced shear associated with pixel $j$, $\epsilon_\textrm{s}$ is a value for the intrinsic ellipticity associated with galaxy shapes randomly sampled from a zero-mean Gaussian distribution with width equal to $\left<\gamma_{\textrm{int}}^2\right>$, and $N_{\textrm{gal}}$ is the number of galaxies that fall within that pixel. 
We produce 200 map realisations of each tomographic power spectrum in this way.
Following this we compute the auto- and cross-correlation power spectra of these maps using the \texttt{PYTHON} package \texttt{HEALPY}\footnote{\href{https://github.com/healpy/healpy}{https://github.com/healpy/healpy}}. We use these power spectra to evaluate the covariance matrix in 14 multipole bins in the range $10 \leq \ell \leq 4,000$, which we calculate as:
\begin{multline}
    \label{eq:cov_mat}
    \textrm{cov}\left[P^{\gamma}_{ij}(\ell), P^{\gamma}_{i'j'}(\ell') \right] = \\
    \left<\left(P^{\gamma}_{ij}(\ell) - \left<P^{\gamma}_{ij}(\ell)\right> \right)
           \left(P^{\gamma}_{i'j'}(\ell') - \left<P^{\gamma}_{i'j'}(\ell')\right> \right)\right>
\end{multline}

The resultant cosmic shear auto- and cross-power spectra can be seen in Fig.~\ref{fig:shear} for all cosmologies, along with the associated uncertainties on the cosmic shear power spectrum, which serves to illustrate the absolute power spectra and how the associated changes due to the different cosmological models are generally quite subtle. The differences in the cosmic shear power spectra are better highlighted in Fig.~\ref{fig:ratio}, which shows the ratio of each auto- and cross-correlation power spectrum with respect to the \LCDM prediction in that same tomographic bin.  We discuss here the calculation of the error bars in Figs. \ref{fig:shear} and \ref{fig:ratio} and leave the scientific interpretation of these results for Section \ref{sec:results}.

The red error bars in Fig.~\ref{fig:ratio} show the theoretical uncertainty associated with the shear power spectrum as computed using equation \ref{eq:fiducal_err}. The black error bars represent the diagonal elements of the covariance matrix for a \LCDM cosmology calculated using multiple realisations of each shear power spectrum as generated by \FLASK.   Note here that the error bars which are shown are normalised to the \LCDM~power spectrum in each tomographic bin, i.e., $\Delta P^{\gamma}_{ij}(\ell)/P^{\gamma}_{ij}(\ell)$.

It can be seen in Fig.~\ref{fig:ratio} when comparing the theoretical error bars shown in red to the black error bars extracted from the covariance matrix computed with the assistance of \FLASK, that the errors are in excellent agreement (as one would expect) for the auto-correlation power spectra.
There is some disagreement at low multipoles, which stems from cosmic variance issues associated with the maps.\footnote{We tested whether the error bars at small multipole were brought into better agreement if one created full sky realisations of the power spectra using \FLASK~rather than a masked version and found that this was indeed the case. Therefore, it is likely that the masking is adding some additional spurious noise to the signal on large angular scales.}
However, importantly, this agreement does not hold for the cross-correlation power spectra, with significant disagreements at large multipoles (small angular scales).  Specifically, the analytic calculation in equation~\ref{eq:fiducal_err} ignores the effects of Poisson noise on uncertainty in the cross-spectrum, but we find there is always a non-negligible Poisson error. %associated with the cross-spectrum due to chance alignments associated with the uncertainty in galaxy shapes.  

In Appendix \ref{sec:additional_error} we use realisations of pure shape noise fields generated with \FLASK~to derive a more accurate and general equation for incorporating the impact of shape noise (Gaussian errors only) on the cross-spectrum, namely:
\begin{multline}
\label{eq:updated_err}
\Delta P_{ij}^{\gamma}(\ell) = \\ 
\sqrt{\frac{2}{\left(2\ell+1\right)f_{\textrm{sky}}}} 
\left[P_{ij}^{\gamma}(\ell) + \delta_{ij} \frac{\left<\gamma_{\textrm{int}}^2\right>}{\overline{n_i}} +
\left(1-\delta_{ij}\right)\frac{\left<\gamma_{\textrm{int}}^2\right>}{\sqrt{2\overline{n_i}\overline{n_j}}}
\right].    
\end{multline}

The new term (right most in the square brackets) represents the contribution of Poisson noise to the uncertainty in the cross-power spectrum.  Note the similarity of the noise terms for the auto- and cross-spectra, which is expected because they are caused by the same effect: random alignments of galaxy shapes.  The only difference is that for the cross-spectrum term we allow for the possibility of different source densities in the tomographic bins being cross-correlated, and the extra factor of square root of 2, which is the result of having more galaxy pairs to evaluate in the cross-spectrum compared to the power spectrum (the error scales as the number of pairs).

Finally, we note that the black error bars plotted in Figs. \ref{fig:shear} and \ref{fig:ratio} correspond to just the diagonal elements of the covariance matrix, but we use the full covariance matrix when evaluating signal-to-noise ratios in  in Section \ref{sec:results}.

\subsection{Summary of theoretical cosmic shear pipeline}
Here we provide a very brief summary of the steps described above in generating cosmic shear power spectra for the different cosmological models in this study, alongside a realistic estimate of the uncertainties.
\begin{itemize}
    \item Compute $P(k)$ for each cosmological simulation (including box size variations for the two $\alpha_s$ simulations) and re-bin to smooth out some of the initial numerical noise associated with the power spectra.
    \item Compute the ratio of each power spectrum with respect to a corresponding $\Lambda$CDM simulation (of same box size and resolution). 
    \item For cosmologies with a running spectral index, combine the ratios computed from the {400, 100, 25} Mpc $h^{-1}$ box simulations along with the \texttt{Halofit} prediction to a $\log(k_{\rm{min}})=-4$. For all other cosmologies, extrapolate $R_{C}(k, z) = 1$ for large scales not sampled in the 25 Mpc $h^{-1}$ box. Fit a cubic spline to the constructed ratios and smooth.
    \item Calculate the absolute $P(k)$ for each cosmology using the constructed ratio as a boost factor to modify the $\Lambda$CDM prediction computed using \texttt{Halofit} (equation \ref{eq:boost_factor}).
    \item Generate theoretical tomographic shear power spectra for each cosmological model using equation \ref{eq:shear_pow}.
    \item Construct multiple synthetic tomographic weak lensing shear maps for a $\Lambda$CDM universe including galaxy shape noise using the \texttt{FLASK} package. Compute $C(\ell)$ for each realisation to construct a covariance matrix for the $\Lambda$CDM prediction using equation \ref{eq:cov_mat}.
\end{itemize}

\section{Results}
\label{sec:results}
In this section we present the main results of this work. This includes predictions for the tomographic cosmic shear auto- and cross-power spectra for each cosmological model, along with their respective ratios with respect to $\Lambda$CDM. We also explore how large the differences in the cosmic shear power spectrum due to a change in cosmology are compared to the expected error associated with the cosmic shear power spectrum.
\subsection{Comparisons up to \texorpdfstring{$\ell_{\textrm{max}} = 4,000$}{lmax=4000}}
\label{sec:l_max_4000}

\begin{figure*}
    \centering
    \includegraphics[width=\textwidth]{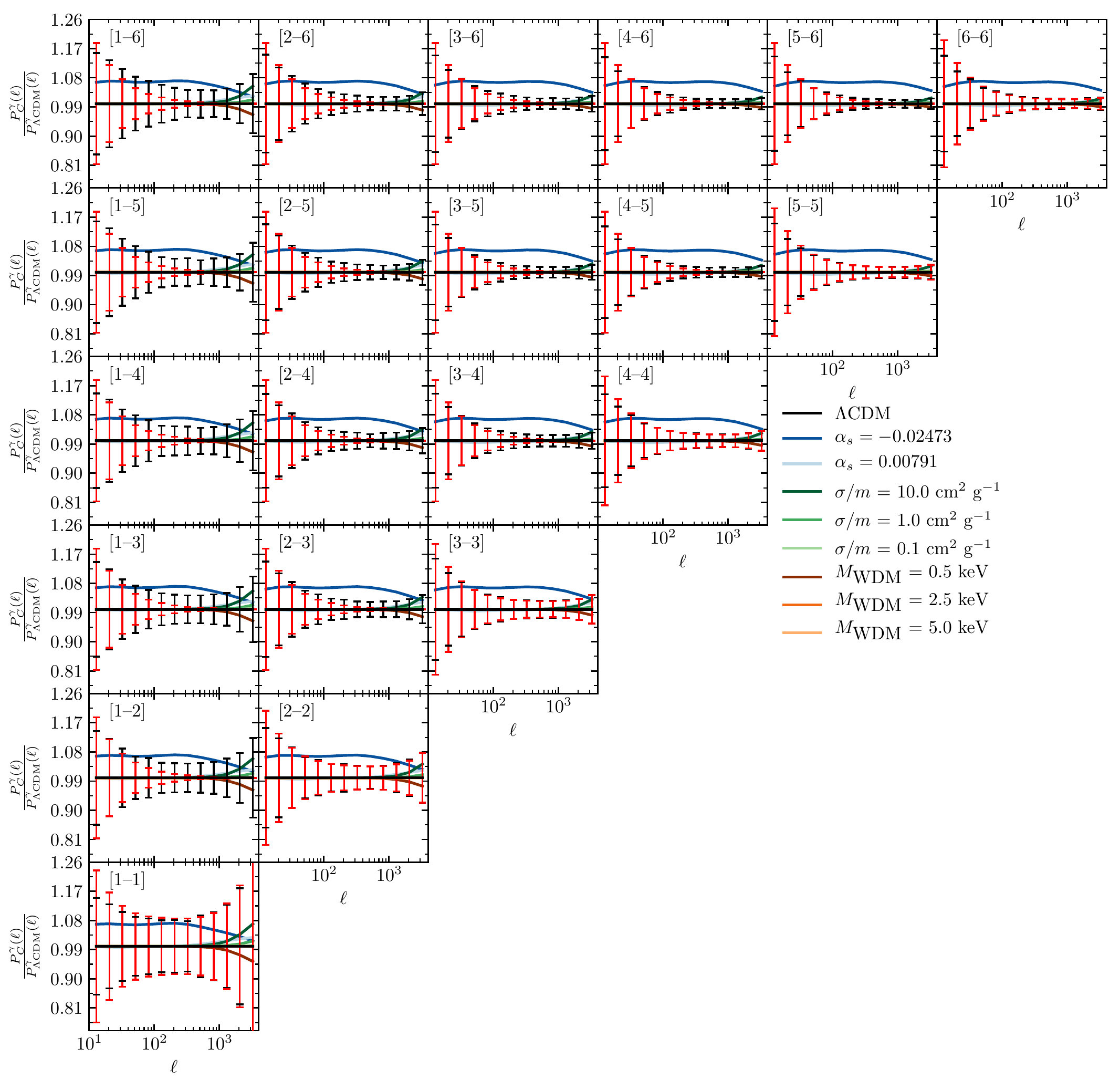}
    \vspace{-0.5cm}
    \caption{The shear auto- and cross-correlation power spectrum for each cosmological model normalised to the corresponding \LCDM auto/cross-correlation power spectrum. The different coloured lines represent the different cosmological modes, indicated in the legend. The red error bars show the standard theoretical (analytic) prediction for the noise on the shear power spectra (which does not include the effects of Poisson noise on the shear cross-spectra), and the black error bars show the noise computed using multiple realisations of each auto and cross power spectrum in a \LCDM cosmology with the \FLASK~package. This plot helps highlight the resultant changes to the cosmic shear power spectrum, particularly due to the inclusion of a running scalar spectral index as a free parameter in the standard model. The numbers in square brackets in the top left of each panel indicate the tomographic bin number. Increasing tomographic bin number corresponds to higher redshifts (see Table \ref{Tab:redshifts} for the mean redshift of each bin).}
    \label{fig:ratio}
\end{figure*}

We now discuss the main results of our investigation for the fiducial \Euclid-like setup, examining the cosmic shear power spectrum up to a maximum multipole of $\ell_{\textrm{max}} = 4,000$.  

Examining the ratios of the various cosmological extensions with respect to $\Lambda$CDM shown in Fig.~\ref{fig:ratio}, we can immediately conclude (by eye) that all of the cosmological extensions studied here are capable of producing some level of deviation in the cosmic shear power spectrum compared to the $\Lambda$CDM prediction over this range of multipoles. However, in the case of WDM and SIDM, these changes appear most noticeable for the two extremest models examined in this study. The other models do produce differences as well, but they are comparatively smaller and a more quantitative analysis of their `detectability' is thus required.

\begin{figure*}
    \centering
    \includegraphics[width=\textwidth]{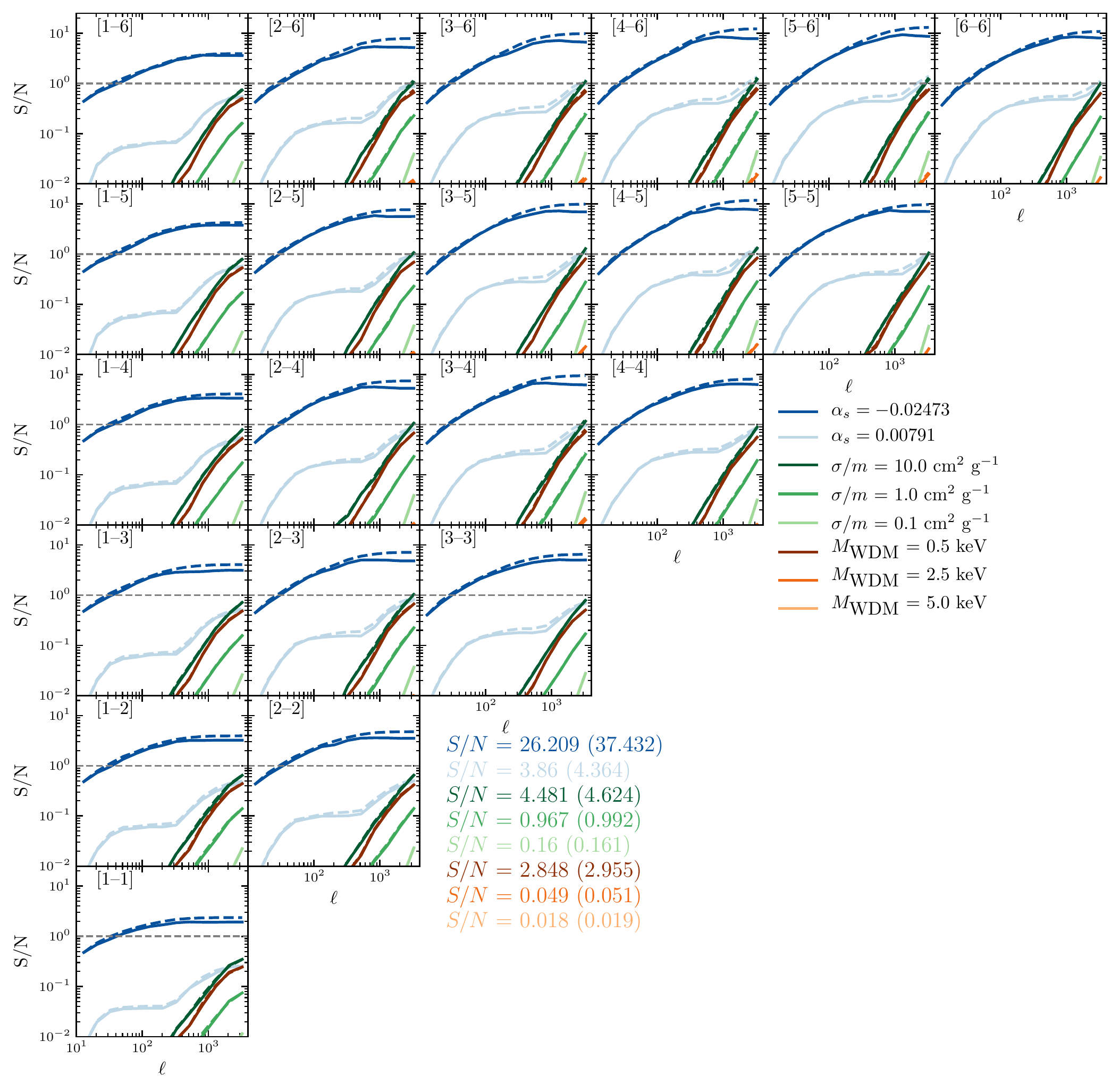}
    \vspace{-0.5cm}
    \caption{The integrated signal to noise ratio (SNR) as a function of $\ell_{\textrm{max}}$ to demonstrate how significant the deviations in the shear power spectra are for the different cosmological models relative to \LCDM. We plot here two variants of this statistic shown by solid and dashed lines. In the first case (solid line), the SNR is calculated using the entire covariance matrix, whereas the dashed line only uses the diagonal elements of the covariance matrix. The total integrated SNR, summed (in quadrature) over the tomographic bins, is shown next to the bottom left panel with the values in brackets indicating the SNR if one only uses the diagonal elements of the covariance matrix. }
    \label{fig:SNR}
\end{figure*}

We quantitatively characterise the constraining power of the cosmic shear observations via a signal-to-noise ratio (SNR), which we evaluate as:

\begin{multline}
\label{eq:SNR}
\left(S/N\right)^2 = \\\sum_{\ell, \ell' \leq \ell_{\textrm{max}}} (\left|R(\ell)-1\right|)\left(\frac{\rm{Cov}[P^{\gamma}_{ij}(\ell), P^{\gamma}_{ij}(\ell')]}{P^{\gamma}_{ij}(\ell) P^{\gamma}_{ij}(\ell')}\right)^{-1}
(\left|R(\ell')-1\right|) \ \ \ ,
\end{multline}

\noindent where $R(\ell)$ is the ratio of each cosmological extension with respect to \LCDM~as plotted in Fig.~\ref{fig:ratio}. We make use of the full covariance matrix calculated using equation \ref{eq:cov_mat}, which we normalise with respect to the absolute power spectra themselves. We do this because we are comparing the error bars to the ratios rather than the absolute power spectra.

The resultant plot of the integrated SNR as a function of $\ell_{\textrm{max}}$ can be seen in Fig.~\ref{fig:SNR}. The solid curves represent the integrated SNR as a function of scale when using the full covariance matrix to estimate the uncertainties.  For comparison, the dashed curves show the integrated SNR when using only the Gaussian (diagonal elements) errors.  A legend is provided which lists the integrated SNR when summed (in quadrature) over all tomographic bins for the full covariance matrix case and for the diagonal errors only (the latter is in parentheses).  

As expected from Fig.~\ref{fig:ratio}, the cosmology with a negative value for the running of the spectral index shows strong deviations from the \LCDM~prediction over almost the entire multipole range, with these differences becoming larger at higher redshift. The reason for this stems mainly due to an increase in the absolute power on these multipoles in the later tomographic bins, as seen in Fig.~\ref{fig:shear}. This causes the relative error bars to decrease with increasing redshift resulting in a larger SNR.  When summed over all tomographic bins, a Planck-based cosmology with a negative running of the scalar spectral index will likely be easily distinguishable from a Planck-based \LCDM~cosmology with a \Euclid-like cosmic shear survey.

One apparent feature in the SNR for the negative running cosmology is the plateau towards higher multipoles. The reason for this behavior is that the signal we see here is dominated by the region of the power spectrum for which a negative running cosmology produces an enhancement in power (due to the increase in $A_s$) relative to $\Lambda$CDM. However, on the largest multipoles the ratio begins to turn over and decrease, where it would eventually cross over the \LCDM~prediction and predict a suppression in the cosmic shear power spectra, rather than the enhancement that we see on these scales. Therefore, the signal, relative to the error bars, decreases over this range of scales yielding a plateau in the integrated SNR.
The same effect is also seen, perhaps somewhat more clearly, in the cosmology with a positively running scalar spectral index where the opposite is true. 
In this case, there is initially a slight suppression in the cosmic shear power spectra which flips and becomes an enhancement at high multipoles, producing the initial increase in the SNR, followed by a plateau and then a second increase after the ratio has crossed unity. 
As a result, the positive running cosmology has an integrated SNR in the later multipole bins exceeding 1, showing that upcoming future surveys such as \Euclid~and \LSST~are perhaps able to put constraints on a value for the running of the scalar spectral index that are competitive to those from cosmic microwave background and Lyman-$\alpha$ forest constraints.

Note that the SNR would likely continue to increase if we increased the maximum source redshift beyond the limit of $z_s$=1.0 that we adopt here, at least for the running cosmologies.
If one was to include tomographic redshift bins beyond $z_\textrm{s}$ = 1.0 the relative error bars shown in Fig.~\ref{fig:ratio} would decrease (due to the increase in the cosmic shear signal); hence this would enhance the SNR seen here.
The reason we do not include these higher redshift tomographic bins in this study is because, as mentioned, at fixed $\ell_{\textrm{max}}$, but increasing $z_\textrm{s}$ the region of the matter power spectrum one becomes sensitive to tends towards smaller $k-$modes. 
As such, although this would still result in a measurable signal for the two models with a running spectral index, it would not result in a measurable signal for the WDM and SIDM cosmologies (their effects are confined to small physical scales).
However, these higher redshift tomographic bins will be included in future weak lensing surveys, resulting in potentially increased constraining power on $\alpha_s$ compared to what is shown here. 

For the SIDM and WDM cosmologies it appears that one needs to probe to much smaller scales (higher multipoles) to be able to distinguish most of the models we explored from $\Lambda$CDM.  The only exceptions are the most extreme WDM (0.5 keV) and SIDM ($\sigma/m = 10$ cm$^2$ g$^-1$) models.  Our calculations suggest that a \Euclid-like survey with realistic source densities and shape noise may be able to (marginally) distinguish these models from $\Lambda$CDM.  Pushing to higher multipoles would help in these cases as well.

An important aspect of a \Euclid-like survey is the fact that it has a large sky coverage, meaning the cosmic variance error associated with the measurements of the cosmic shear power spectrum is strongly reduced on all but the largest angular scales. 
Therefore, the limiting factor in the ability of detecting a difference between some of these cosmological models and $\Lambda$CDM, particularly at the upper-end of the multipole range we analyse in this study, is a result of the uncertainty in galaxy shapes. 
Thus, we have also investigated a similar tomographic setup as that expected for the \NGRST~survey which instead of a large $f_{\textrm{sky}}$ (\NGRST~is envisioned to have $f_{\textrm{sky}}$ = 0.0485), is expected to have a larger number density of source galaxies equal to $n_{s}$ = 51 galaxies/arcmin$^2$ compared to $n_s = 30$ galaxies/arcmin$^2$ for a \Euclid-like (or \LSST-like) setup. 

For the \NGRST~setup we use five tomographic bins up to redshift 1.0, defining the source galaxy distribution using equation \ref{eq:n_z} (here we adopt the parameter values: $\alpha = 2$, $\beta = 0.9$, $z_0 = 0.28$ in order to closely mimic the galaxy distribution described in \citealt{Eifler2020}). 
We find that, although the shot noise error is decreased in such a survey setup, the increase in the cosmic variance error due to the reduction in sky coverage compensates, resulting in no improvement on the SNR\footnote{The total integrated SNRs calculated for a \NGRST-like setup for each cosmology, using the full covariance matrix are: \{9.657, 1.733, 2.203, 0.481, 0.082, 1.39, 0.023, 0.009\}, in the same order as that displayed in Fig.~\ref{fig:SNR}.}.
Therefore, it seems that one would need both a large $f_{\textrm{sky}}$ (although potentially not as large as \Euclid~or \LSST), and a large number density of source galaxies, $n_s$, to use cosmic shear to better constrain these cosmological models without having to push to higher multipoles. 

It is apparent from the above analysis that, if it was possible to push to higher multipoles, this could considerably increase the constraining power for the SIDM and WDM scenarios.  
However, if we were to evaluate the uncertainties using a full covariance matrix approach this would be much more computationally expensive, as each realisation has around $2\times10^8$ pixels in each tomographic map and ($N_{\textrm{T}}(N_{\textrm{T}}+1)/2$) auto- and cross-correlation power spectra (where $N_{\textrm{T}}$ is the number of tomographic bins)\footnote{Note that we produce 200 unique realisations in order to evaluate the covariance matrix.}.  
However, we can considerably simplify the process if the Gaussian errors (in equation~\ref{eq:updated_err}) are sufficient to calculate a SNR.  Comparing the solid and dashed curves in Fig.~\ref{fig:SNR}, we conclude that the difference in the integrated SNRs are fairly significant for the running cases (particularly the negative running scenario), but not for the SIDM or WDM cases, at least over the range of multipoles examined here.  As such, below we examine if pushing to $\ell_{\textrm{max}} = 20,000$ improves the SNR for the models with WDM, SIDM and positive running scalar spectral indices, using the estimated Gaussian errors rather than a full covariance matrix approach.

\subsection{Comparisons up to \texorpdfstring{$\ell_{\textrm{max}} = 20,000$}{lmax=20000}}
\label{sec:l_max_20000}

\begin{figure*}
    \centering
        \includegraphics[width=\textwidth]{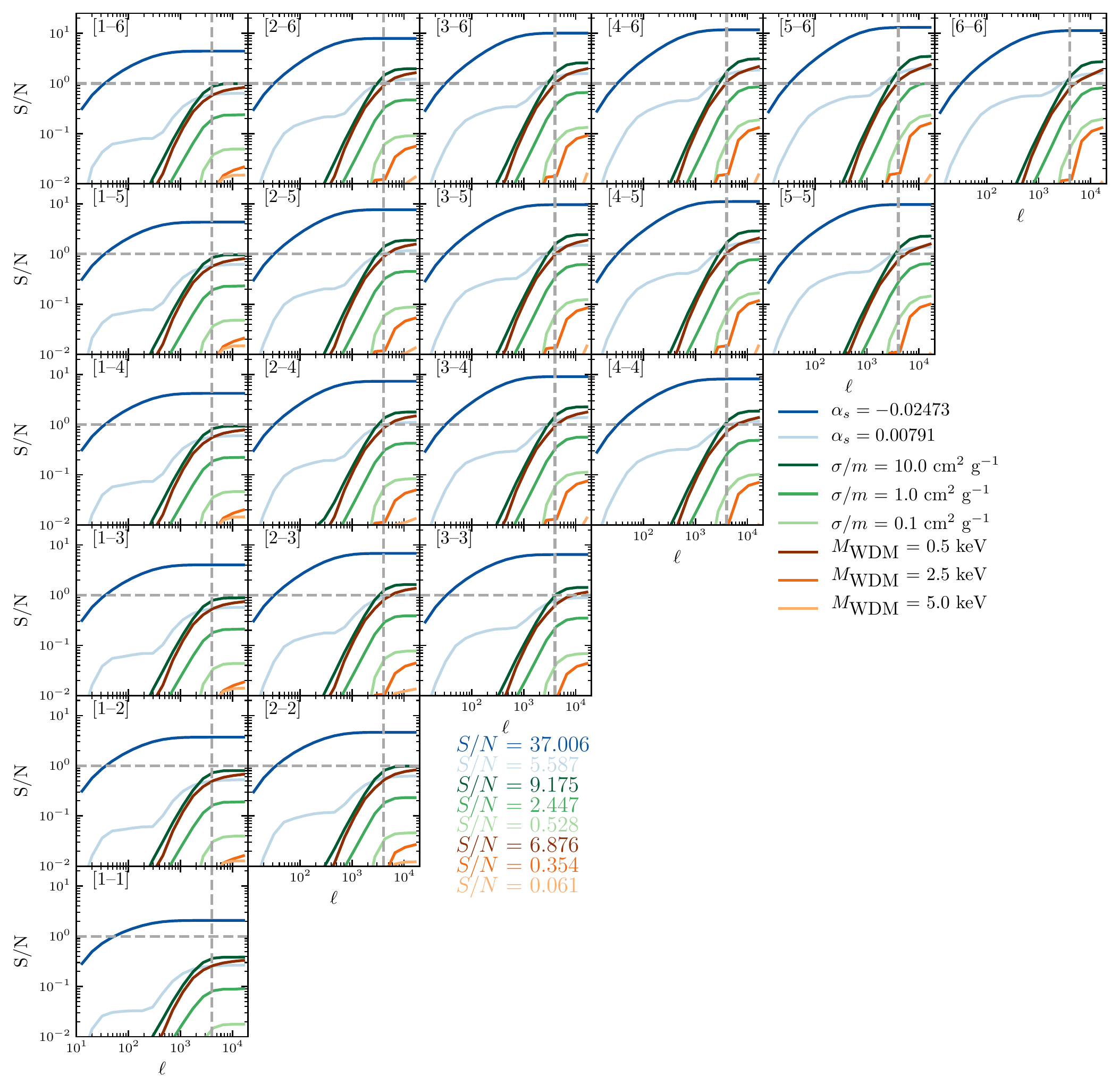}
    \vspace{-0.5cm}
    \caption{Same as Fig.~\ref{fig:SNR}, however, here we plot the integrated SNR up to an increased $\ell_{\textrm{max}}$ of 20,000. To do this we only make use of the Gaussian terms of the covariance matrix, computed using the updated error equation shown by equation \ref{eq:updated_err}. This shows that if one was able to push to these smaller angular scales, these observations could potentially be fruitful in putting constraints on the the cross-section for interaction of dark matter particles, as well as increased constraining power on the running of the spectral index. The vertical grey dashed line corresponds to the previous $\ell_{\textrm{max}}$ = 4,000.}
    \label{fig:extended_SNR}
\end{figure*}

We show in Fig.~\ref{fig:extended_SNR} the new integrated SNR, extended up to an $l_{\textrm{max}}$ = 20,000, with the vertical dashed grey line in each panel showing the previous $l_{\textrm{max}}$.
Note that we have replaced the reduced covariance matrix term in equation \ref{eq:SNR}, with simply $\left(\Delta P_{ij}^{\gamma}/P_{ij}^{\gamma}\right)^{-2}$, where $\Delta P_{ij}^{\gamma}$ is calculated using equation \ref{eq:updated_err}, which includes the additional Poisson error on the cross-correlation power spectra.

Pushing to higher multipoles does indeed reveal regions of the cosmic shear power spectrum where the differences due to changes in cosmology are not yet completely drowned out by the associated noise.  This is particularly true for the positive running cosmology and the two SIDM cosmologies with the smallest cross-sections, which all show a marked improvement in the total integrated SNR when summed over tomographic bins, as shown at the bottom of Fig.~\ref{fig:extended_SNR}. 
However, although the total integrated SNR is now > 1 for the less extreme SIDM cosmology of the two, the integrated SNR never exceeds 1 in any single tomographic bin, meaning the full tomographic information will be needed to place constraints on the cross-section for interaction. 

Note that the signal present in each tomographic bin will be somewhat sensitive to the tomographic binning strategy.  For example, at fixed source density for a given survey, the fewer tomographic bins one has the larger the effective number density of source galaxies in each tomographic bin.
This will work to decrease the shot noise associated with each tomographic power spectrum, thus increasing the integrated SNR in an individual bin.  The integrated SNR when summed over all bins, though, should be a more robust quantity.

In the case of the WDM cosmologies, pushing to these higher multipoles does not result in a significant gain in SNR, insofar as detection is concerned.  As explained previously, this is in part because the suppression in the matter power spectrum in a WDM cosmology is maximal at larger redshifts.  However, the $k-$scales affected in a WDM cosmology move out of range of the $\ell-$modes covered in this study at higher redshifts. 

While it is clear from the above analysis that pushing to higher multipoles in general results in increased SNRs, a potentially important caveat is that our analysis does not take into account additional sources of uncertainty that may be prevalent on such small angular scales, including source deblending difficulties. 
This refers to the phenomenon where sources (be them galaxies or stars) overlap on the sky, disrupting the shear estimation of a source galaxy.
This can be accounted for by rejecting objects which are flagged as blended.
However, as was shown in, for example \cite{Hartlap2011} and \cite{MacCrann2017}, this can lead to a selection bias on the source galaxies used in the cosmic shear analysis.  In particular, this selection bias works to exclude galaxies in high density environments, which will have a higher convergence than average, resulting in a biased (low) two-point correlation function of galaxy shapes, particularly on small-scales (see e.g., fig.8 in \citealt{MacCrann2017}). However, recently \cite{Hoekstra2021} showed that this bias can be effectively mitigated using a process of \texttt{METADETECTION} \citep{Sheldon2020}.
In addition to this, an effective modelling of baryonic physics will be needed on these scales to fully extract the cosmological information on these small scales (see below).

\section{Discussion \& Conclusions}
\label{sec:conclusions}

In this study we have explored the effects that different extensions to the standard model of cosmology have on the non-linear matter power spectrum, particularly on small scales.  This was achieved using a suite of numerical simulations which contain three cosmological variations (in addition to the fiducial $\Lambda$CDM realisation): i) a running scalar spectral index ($\alpha_s$), warm dark matter ($M_{\textrm{WDM}}$) and self-interacting dark matter ($\sigma/m$).  We focus on these extensions in particular as they have previously been shown to suppress small-scale structure and therefore offer a potential means to mitigate small-scale challenges which have been highlighted with the $\Lambda$CDM model.
We combined the small-scale power spectra extracted from the simulations with the predictions of \texttt{Halofit} on large scales to construct non-linear matter power spectra for the different extensions spanning over six orders of magnitude in wavenumber ($-4 \leq \log_{10}(k [h\textrm{Mpc}^{-1}]) \leq 2.4$). 
These power spectra were used to compute the cosmic shear power spectrum.
Finally, We evaluated via synthetic lensing observations (generated with \FLASK) whether forthcoming Stage-IV lensing surveys (\Euclid, \LSST, and \NGRST) will potentially be able to differentiate these extensions from the standard $\Lambda$CDM model.

The main findings of our study are as follows: 
\begin{itemize}
    \item A negative running spectral index, WDM and SIDM are all capable of producing a significant suppression in the non-linear matter power spectrum at late times (Fig.~\ref{fig:non_linear_Pk}). 
    At $z = 0$, this suppression can range from $\approx5\%$ to $\approx40\%$ in the case of WDM, at $k \approx$ 100 \hpMpc. 
    Furthermore, in the case of WDM, the suppression in the matter power spectrum increases  with increasing redshift, rising to $\ga 60\%$ at $z = 1$.
    A similar trend is seen for the negative running cosmology, although not to the same extent as for WDM. Conversely, the suppression in the matter power spectrum increases with decreasing redshift in an SIDM cosmology. This is due to the strong cores developing in the density profiles of haloes as structures collapse and the relative velocities of particles increase. 
    \item From the different cosmological extensions we have examined, a running scalar spectral index looks the most promising in having a measurable effect on the cosmic shear power spectrum for upcoming surveys such as \Euclid. This can be seen, somewhat counter-intuitively, in the enhancement that is produced in the cosmic shear signal on intermediate scales, due to the change in the amplitude of the non-linear power spectrum with respect to $\Lambda$CDM on the important scales (Fig.~\ref{fig:ratio}). We find that there is a significant signal for both a negatively running spectral index and a positively running one, shown through the signal-to-noise ratio (Fig.~\ref{fig:SNR}). This illustrates that cosmic shear could be an additional probe which, if used in combination with other probes such as the CMB, could help place strong constraints on the running $\alpha_s$.
    \item The constraining power for the other two cosmological extensions of interest here, the mass of the WDM particle and the self-interaction cross-section, is slightly weaker, owing to the fact that these extensions affect only the smallest scales.  We have shown that upcoming cosmic shear measurements should be able to rule out SIDM models with $\sigma/m$ > 10 cm$^2$ g$^{-1}$ or WDM models with thermal relic masses $M_{\textrm{WDM}}$ < 0.5 keV.  While these are perhaps weaker constraints than what can be obtained from other methods (e.g., satellite abundances, strong lensing time delay, Lyman-$\alpha$ forest), we note that cosmic shear is independent test that has very different systematics than previous small-scale probes, making it still a very worthwhile test of these small-scale extensions.  In addition, we have demonstrated that, if it is possible to push to higher multipoles with these experiments, there is the potential that cosmic shear could help place some of the strongest constraints on the SIDM cross-section ($\sigma/m$) (Fig. \ref{fig:extended_SNR}). 
    \item Finally, we have illustrated that the standard analytic prediction for the error associated with the cosmic shear cross-correlation power spectrum (between tomographic bins) significantly underestimates the true error that one recovers when computing the same power spectrum from a map which includes a prescription for galaxy shape noise by a factor of around 20 at $\ell \approx 1,000$ (Fig.~\ref{fig:err_ratio}) .  This is because the standard prescription ignores the associated Poisson error.  We have introduced a modification to the analytic form of the error which modifies the error associated with cross-correlation power spectrum (equation \ref{eq:updated_err}) and is found to bring the analytic errors into much better agreement with our empirical findings based on cross-correlating shape noise maps.
\end{itemize}

One of the main focuses of upcoming weak lensing surveys, such as \Euclid, \LSST~and~\NGRST, is to help place constraints on the dark energy equation of state parameter $w$, or the time varying dark energy equation of state parameters $w_0, w_a$.  It is well-established that cosmic shear tomography provides a sensitive probe of the growth of structure which, in turn, depends on the evolution of dark energy.
However, here we illustrate that the constraining power of cosmic shear measurements also extends beyond dark energy and the other Friedmann parameters.  Specifically, we have shown that cosmic shear observations can potentially place constraints on the running of the spectral index, as well as the cross-section for interaction of dark matter particles (SIDM) and the thermal relic mass (WDM).
As a final test, we have calculated the source density required to improve constraints on the less extreme cosmological models such as the $M_{\textrm{WDM}} = 2.5, 5.0$ and $\sigma/m = 0.1$ cm$^2$ g$^{-1}$ models. We find that for a sky coverage equal to that of \Euclid, the required source density of background galaxies should be $\approx$ \{85, 525, 60\} galaxies/arcmin$^{2}$, respectively, in order to obtain a SNR of > 1, when pushing the analysis to $\ell_{\textrm{max}}$=20,000. 

The tests performed in this study consisted of whether forthcoming cosmic shear surveys could distinguish (on the basis of SNR) a number of extensions, with specific parameter values, from the baseline $\Lambda$CDM model.  The tests were particularly simple in that, in most cases, all of the cosmological parameters were held fixed, apart from the new parameters describing the extension.  A more realistic test would be to allow the various parameters to be free and to marginalise over them when estimating the uncertainty (and potentially bias) in the recovered extension parameters (e.g., WDM particle mass). However, to achieve this requires many more simulations than produced here and potentially sophisticated methods (e.g., emulators) for interpolating the results for arbitrary cosmological parameter values.  While our promising results demonstrate that this is clearly worthwhile, it is a large undertaking and we leave this for future work.

In this study we focused on the constraining power of future cosmic shear measurements. However, there are additional complementary two-point statistics that may be helpful in placing constraints on these cosmological extensions. In particular there is galaxy clustering, which describes the clustering between lens galaxies, and galaxy--galaxy lensing, which describes the over-density of mass around lens galaxies. Together with cosmic shear, these 3$\times$2pt statistics have been shown to help place tighter constraints on the cosmological parameters $\Omega_{\textrm{m}}$ and $\sigma_8$ (see e.g. fig. 6 of \citealt{DES2021results}). However, combining all three of these probes is beyond the scope of this work, which focused on the constraining power of cosmic shear alone. In future work, we will explore how the constraints in this paper change when combining these 3$\times$2pt statistics.

An important caveat to the work we have presented is that the simulations we used neglected the effects of baryonic physics.  
Previous work has demonstrated that when numerical simulations include complex galaxy formation physics, such as feedback from supernovae and active galactic nuclei, they can produce relatively large effects (typically 5-20\%) on the non-linear matter power spectrum (e.g., \citealt{vanDaalen2011,Chisari2018,Springel2018,vanDaalen2020}).
Furthermore, these effects due to galaxy formation physics have been shown to be visible in the cosmic shear two-point correlation functions at a similar level \citep{Semboloni2011}.
Our work has focused mostly on small scales where, as opposed to baryons suppressing the power spectrum via the expulsive effects of feedback on relatively large scales (the focus of most previous studies), it is more likely that cooling and star formation will lead to an \textit{enhancement} in the power spectrum.  Regardless of whether baryons produce a suppression or an enhancement, the effects may be degenerate with the cosmological extensions we have examined here.  As discussed in \citet{Stafford2020b}, ultimately what is required is a systematic and simultaneous exploration of the (uncertain) impact of baryons and cosmological variations on small scales and an understanding of how these effects propagate through to observables on small scales such as the cosmic shear power spectrum. 

In closing, to help mitigate potential degeneracies, recent studies have shown the power of combining complementary probes when placing constraints on additional cosmological (and baryon) parameters. For example, \cite{Enzi2020} and \cite{Nadler2021} have illustrated the power of combining multiple probes (Lyman-$\alpha$ forest, strong lensing and the abundance of Milky Way satellites) in placing constraints on the WDM particle mass.  Measurements of cosmic shear on small scales provide an important new tool in this regard and one that has very different systematic uncertainties from currently used small-scale probes.  Forthcoming Stage-IV lensing surveys therefore offer a promising new window to study cosmological and galaxy formation physics on small scales.

\section*{Acknowledgements}

The authors warmly thank Henk Hoekstra for helpful feedback on this paper.  SGS acknowledges an STFC doctoral studentship. This project has received funding from the European Research Council (ERC) under the European Union's Horizon 2020 research and innovation programme (grant agreement No 769130). AR is supported  by the European Research Council's Horizon 2020 project `EWC' (award AMD-776247-6). This work used the DiRAC@Durham facility managed by the Institute for Computational Cosmology on behalf of the STFC DiRAC HPC Facility. The equipment was funded by BEIS capital funding via STFC capital grants ST/P002293/1, ST/R002371/1 and ST/S002502/1, Durham University and STFC operations grant ST/R000832/1. DiRAC is part of the National e-Infrastructure.

Some of the results in this paper have been derived using the \texttt{HEALPY} and HEALPix package.

\section*{Data Availability Statement}
The data underlying this article will be shared on reasonable request to the corresponding author.

%%%%%%%%%%%%%%%%%%%%%%%%%%%%%%%%%%%%%%%%%%%%%%%%%%



\bibliographystyle{mnras}
\bibliography{references} % if your bibtex file is called example.bib

%%%%%%%%%%%%%%%%%%%%%%%%%%%%%%%%%%%%%%%%%%%%%%%%%%

%%%%%%%%%%%%%%%%% APPENDICES %%%%%%%%%%%%%%%%%%%%%
\appendix

\section{Resolution and box size study}
\label{sec:resolution_test}

\begin{figure*}
    \centering
    \includegraphics[width=\textwidth]{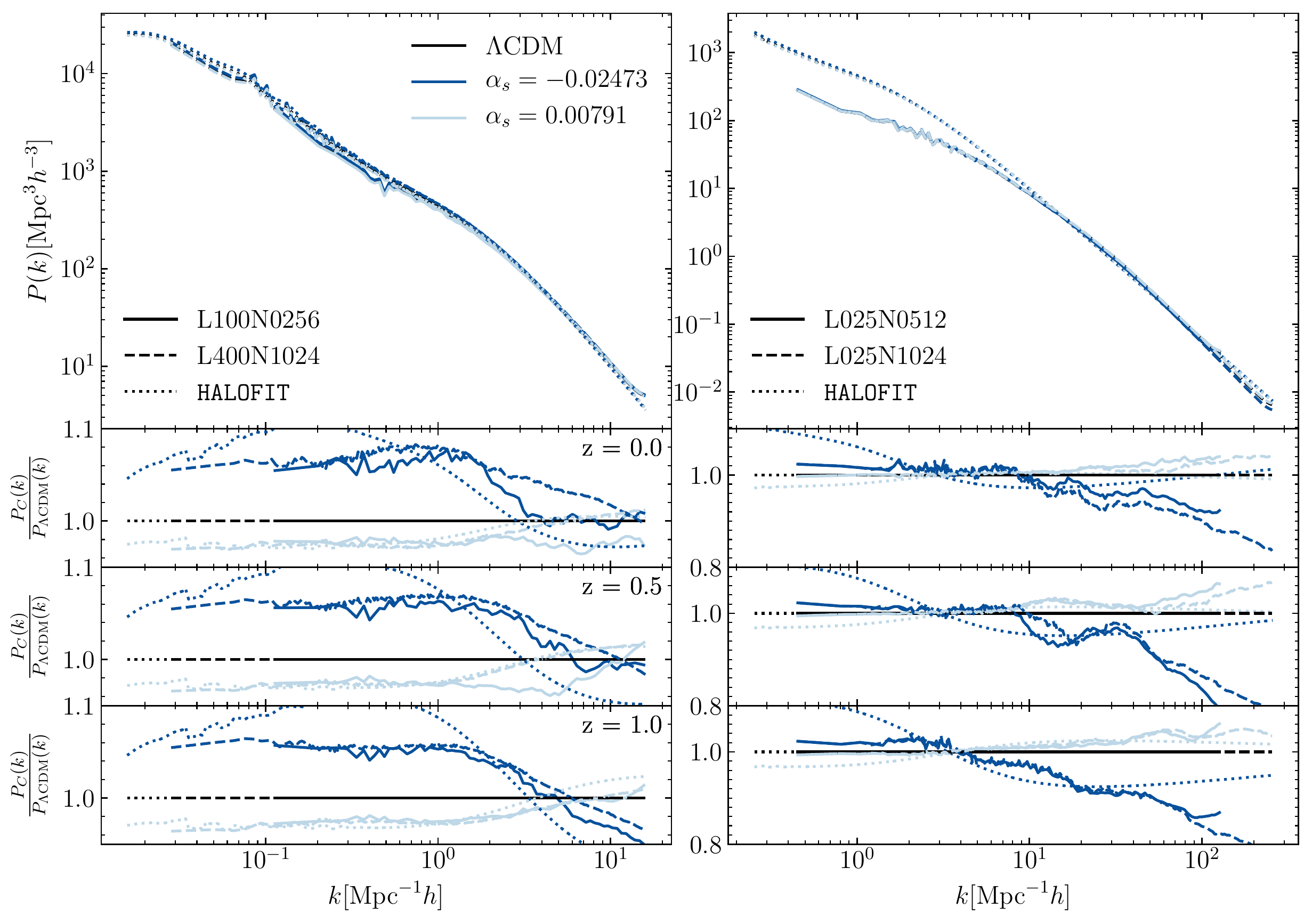}
    \vspace{-0.5cm}
    \caption{Testing how both box size and resolution affect the ratios of the power spectrum extracted from a cosmology with a running spectral index with respect to the \LCDM result. Left panels show the results for fixed resolution and varying box size, while the right panels show the opposite. The top panel in each case shows the absolute power spectra at $z=0$. The three sub-panels below show the ratio as a function of redshift.
    The ratio is computed with respect to the power spectrum from the $\Lambda$CDM simulation with the same box size and resolution as in the running spectral index case. The ratio of the power spectra appears to be well converged for the various box sizes and resolutions examined here which motivates us to combine the ratios over the extended $k-$scale, as described in Section \ref{sec:constructing_mat_pow}, rather than the absolute power spectra.}
    \label{fig:res_test}
\end{figure*}

In Section \ref{sec:constructing_mat_pow} we detail our method for constructing the non-linear matter power spectrum spanning a large range of $k$-modes, which is done using the ratio of the matter power spectrum extracted from the numerical 
simulations with respect to the result from the $\Lambda$CDM simulations.  As we show here, splicing ratios of power spectra, rather than absolute power spectra, is more robust to changes in resolution and box size.  We do this using an additional sub-suite of simulations, examining the running spectral index cosmological extensions, which consists of various box sizes and resolutions. Firstly, there are two setups which are used in the analysis detailed in the paper, which have box sizes of 25 and 400 \Mpcph, each with 1024$^3$ particles. We then compare the non-linear matter power spectrum extracted from these simulations to that extracted from simulations which are a step down in resolution. These simulations are 25 and 100 \Mpcph~
on a side and have 512$^3$ and 256$^3$ particles respectively. Thus, this allows us to explore the effects of both box size and resolution on both the absolute matter power spectrum and the ratio with respect to $\Lambda$CDM.

Fig.~\ref{fig:res_test} shows the non-linear matter power spectrum for two running of the spectral index models. In the left panel we show how the simulation box size affects the absolute power spectra (top) and the ratio with respect to a complementary $\Lambda$CDM simulation of the same box size (bottom), both at fixed resolution. In the right panel, we show how resolution affects these two statistics at fixed box size.  Therefore, it is important to note that when comparing the solid to dashed lines (or dotted) in the bottom panels, both the numerator and denominator have changed. Comparing the different linestyles allows us to assess the convergence of the ratio of power spectra from the running cosmologies with respect to a $\Lambda$CDM power spectrum as a function of varying box size at fixed resolution, or vice versa.

One can see that while the absolute power spectra tend to disagree with each other on the largest scales sampled in the box, the ratio is a much better converged quantity. This is particularly relevant for the smaller volume boxes, where the size of the simulations cause an almost order of magnitude suppression in the matter power spectrum on the largest scales sampled by the simulated volume (comparing the simulation curves to the \texttt{HALOFIT} curve). Conversely, if we compare the ratios at a $k-$scale of around 10 \hpMpc, one can see that these agree to within a few percent across the varying box sizes and resolutions.

\section{Analytic error on cross power spectra}
\label{sec:additional_error}
\begin{figure}
    \centering
    \includegraphics[width=\columnwidth]{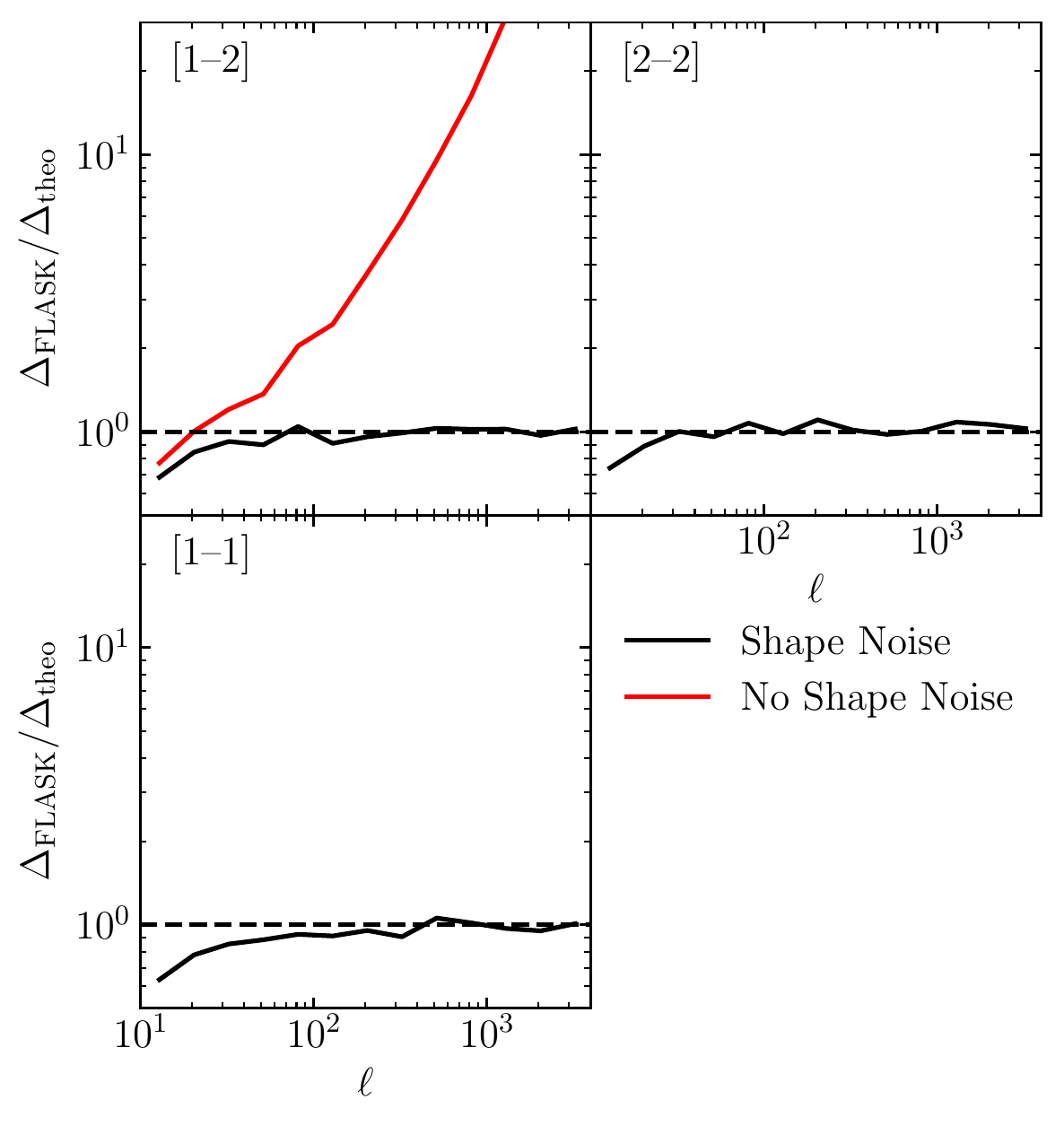}
    \vspace{-0.5cm}
    \caption{The ratio of the errors extracted directly from the multiple realisations of a given \LCDM auto/cross-correlation power spectrum as computed by \FLASK~with respect to the theoretical error bars computed using equations \ref{eq:fiducal_err} \& \ref{eq:updated_err}. Note the tomographic bins shown here are the same as those in the previous plots, however, we do not show the full tomographic setup for brevity as we only need to focus on one cross-correlation power spectrum (with the result being the same for the rest). The black line shows the result when one includes a prescription for the error associated with Poisson noise in the theoretical calculation of the error on the auto/cross-correlation power spectrum, i.e. computed using equation \ref{eq:updated_err}, with the red line being the result if one does not, i.e. computed using equation \ref{eq:fiducal_err}.}
    \label{fig:err_ratio}
\end{figure}

As discussed in Section \ref{sec:theory}, there are two sources of error associated with measurements of the cosmic shear power spectrum (ignoring other systematic errors which exist such as intrinsic alignment errors), these being cosmic variance and Poisson noise. 
It is commonplace to assume that Poisson noise only affects the uncertainty in the auto-correlation power spectrum. However, as we show here, cross-power spectra can still have a significant Poisson noise term.

As shown in Fig.~\ref{fig:ratio}, we find there to be a significant difference in the error bars on small angular scales (large multipoles) when comparing equation~\ref{eq:fiducal_err} to the error bars we derive from synthetic weak lensing maps using \FLASK.  The level of disagreement between the two sets of error bars is more clearly illustrated in Fig.~\ref{fig:err_ratio}, which shows the ratio of the errors extracted directly from the \FLASK~power spectra to the errors computed using equation \ref{eq:fiducal_err}. Note that for clarity we only show here the result for the bottom three panels of Fig.~\ref{fig:ratio}, although the results presented here is true for the other auto- and cross-correlation spectra.
This shows how well the two sets of errors agree with one another in the case of the auto-correlation power spectra, whereas, examining the red curve in the cross-correlation power spectrum panel, the error bars diverge significantly at high multipoles, due to equation \ref{eq:fiducal_err} having no treatment for Poisson noise on the cross-correlation power spectrum.

The fiducial error associated with the cosmic shear power spectrum (shown in equation \ref{eq:fiducal_err}) is unable to capture the residual error which exists on the cross-correlation power spectrum.
This motivates us to produce a more general formula which is able to capture this residual error without having to rerun \FLASK~for each possible tomographic setup.
For this reason, we have run a set of \FLASK~noise-only realisations (with the weak lensing signal due to gravity removed) to derive a functional form for the residual Poisson noise error on the cross-correlation power spectrum. 
The maps were produced using the same tomographic setup described in Section \ref{sec:l_max_4000}, for varying levels of the source density of galaxies.
In total we produced four sets of realisations sampling: $n_s = 15, 30, 45, 60$ galaxies/arcmin$^2$, with each set having 100 realisations of each tomographic bin.
We then computed the auto- and cross-correlation power spectra of these maps, along with cross-correlation power spectra of maps with varying source densities (to test the case where two tomographic bins may not have the same effective number density of galaxies).

We find that an additional term, which only contributes to the cross-correlation power spectrum, is required.  See equation \ref{eq:updated_err}.  This term includes a multiplicative combination of the two effective source densities in each tomographic bin and is able to reproduce the additional error on the cross exceptionally well.
This is illustrated again in Fig.~\ref{fig:err_ratio} in the cross-correlation power spectrum panel (top left), where the black line now shows the ratio of the theoretical error bars, now computed using equation \ref{eq:updated_err}, with the error bars extracted from \FLASK.  Note that here we show the results from the full analysis (i.e., including an intrinsic lensing signal, as well as the noise).  Note that the slight disagreement which exists between the errors on large angular scales is strongly dependent on the sky coverage of the survey. We computed this same test where we had a \Euclid-like galaxy sample, but a full sky survey and this brought the errors into excellent agreement even at low multipoles.

%%%%%%%%%%%%%%%%%%%%%%%%%%%%%%%%%%%%%%%%%%%%%%%%%%

% Don't change these lines
\bsp	% typesetting comment
\label{lastpage}
\end{document}